\documentclass{bmcart}

\usepackage[utf8]{inputenc}
\usepackage{graphicx, url, csquotes}
\usepackage{xcolor,colortbl}
\usepackage{todonotes}
\usepackage{xurl}

\startlocaldefs
\endlocaldefs

\begin{document}

\begin{frontmatter}

\begin{fmbox}
\dochead{Research}


\title{MP Twitter Engagement and Abuse Post-first COVID-19 Lockdown in the UK: White Paper}


\author[
   addressref={aff2,aff1},                   
   email={tracie.farrell@open.ac.uk}   
]{\inits{TF}\fnm{Tracie} \snm{Farrell}}
\author[
   addressref={aff1},
   email={M.E.Bakir@sheffield}
]{\inits{MB}\fnm{Mehmet} \snm{Bakir}}
\author[
   addressref={aff1},
      corref={aff1},   
   email={k.bontcheva@sheffield.ac.uk} 
]{\inits{KL}\fnm{Kalina} \snm{Bontcheva}}


\address[id=aff1]{
  \orgname{Department of Computer Science, Sheffield University}, 
  \street{Regent Court, 211 Portobello},                     %
  \city{Sheffield},                              
  \cny{UK}                                    
}
\address[id=aff2]{
  \orgname{Knowledge Media Institute, The Open University}, 
  \street{Walton Hall},                     %
  \city{Milton Keynes},                              
  \cny{UK}                                    
}


\begin{artnotes}
\end{artnotes}

\end{fmbox}


\begin{abstractbox}

\begin{abstract} 

The UK has had a volatile political environment for some years now, with Brexit and leadership crises marking the past five years. With this work, we wanted to understand more about how the global health emergency, COVID-19, influences the amount, type or topics of abuse that UK politicians receive when engaging with the public. This work covers the period of June - December 2020 and analyses Twitter abuse in replies to UK MPs. This work is a follow-up from our analysis of online abuse during the first four months of the COVID-19 pandemic in the UK. The paper examines overall abuse levels during this new seven month period, analyses reactions to members of different political parties and the UK government, and the relationship between online abuse and topics such as Brexit, government’s COVID-19 response and policies, and social issues. In addition, we have also examined the presence of conspiracy theories posted in abusive replies to MPs during the period.  We have found that abuse levels toward UK MPs were at an all-time high in December 2020 (5.4\% of all reply tweets sent to MPs). This is almost 1\% higher that the two months preceding the General Election. In a departure from the trend seen in the first four months of the pandemic, MPs from the Tory party received the highest percentage of abusive replies from July 2020 onward, which stays above 5\% starting from September 2020 onward, as the COVID-19 crisis deepened and the Brexit negotiations with the EU started nearing completion.

\end{abstract}




\end{abstractbox}
%

\end{frontmatter}



\section{Introduction}

Our previous work studying online abuse in the context of British politics has shown that it can be specific to context, specific to individuals (their characteristics and behaviour) and specific to events unfolding around us \cite{gorrell2020politicians}. Already in the midst of serious upheavals to ``business as usual'' with Brexit and leadership crises, we wanted to understand more about how the global health emergency, COVID-19, influences the amount, type or topics of abuse that UK politicians receive when engaging with the public. 

This white paper charts Twitter abuse in replies to UK MPs between June and December 2020, which is a follow-up from our analysis of online abuse during the first four months of the COVID-19 pandemic in the UK \cite{farrell2020vindication}. The paper examines overall abuse levels during this new seven month period, analyses reactions to members of different political parties and the UK government, and the relationship between online abuse and topics such as Brexit, government's COVID-19 response and policies, and social issues. In addition, we have also examined the presence of conspiracy theories posted in abusive replies to MPs during the period.

This paper makes a contribution to the longitudinal comparison of abuse trends toward UK politicians. Since the same data collection and abuse detection method was used to analyse previous levels of abuse towards MPs in the run-up to the 2017 and 2019 UK General Elections \cite{gorrell2020politicians} and during the first four months of the COVID-19 pandemic in the UK \cite{farrell2020vindication}, this research not only presents new findings, but is also able to corroborate findings of our own previous studies an other related studies. 

Our key new findings are as follows:
\begin{itemize}
\item Abuse levels towards UK MPs in the run up to Brexit in December 2020 reached 5.4\% of all reply tweets sent to MPs. This is the highest level seen across all time periods that we have studied - the 2017 and 2019 General Elections and the first 4 months of the pandemic (Feb -  May 2020).  
\item The 5.4\% average abuse in Dec 2020 is almost 1\% higher than the 4.5\% average abuse levels reached in the two months preceding the 2019 General Election.  
\item Another flashpoint was in October 2020, when abuse levels spiked to almost 5.1\%. Our analysis links this to a specific conflict regarding two MPs and their supporters, however this period also included new tier restrictions, circuit breakers and lockdown protests.
\item In a departure from the trend seen in the first four months of the pandemic, MPs from the Tory party received the highest percentage of abusive replies from July 2020 onwards, which stays above 5\% starting from September 2020 onwards, as the COVID-19 crisis deepened and the Brexit negotiations with the EU started nearing completion. 
\end{itemize}

\section{Related Work}
In this paper, we examine the impact of COVID-19 on abuse levels toward UK MPs. We were expecting impact to be significant, given the amount of misinformation, partisanship and frustration around COVID-19, as well as the existing political affairs of the UK regarding Brexit and party leadership. In a special issue related to online harm during COVID-19, editors Ferrara, Cresci and Luceri~\cite{ferrara2020misinformation} comment that COVID-19 has been an ``unprecedented setting for the spread of online misinformation, manipulation, and abuse, with the potential to cause dramatic real-world consequences''. 

Our previous work, however, was inconclusive about the overall impact of COVID-19 on abuse levels towards UK MPs, due to the novelty of the situation and compassion during Boris Johnson's illness~\cite{farrell2020vindication}. Abuse toward politicians was at an all-time low during Johnson's illness, as he usually features quite prominently in the data because of his role~\cite{gorrell2020politicians}. It is therefore necessary to compare these findings with those of the current period, as the pandemic has matured and Brexit was clearly on the horizon, to see how abuse has levelled-out during this first year of COVID-19. 

Previous work on abuse directed at UK MPs indicated that hostility toward MPs was rising~\cite{gorrell2018twits,gorrell2019race,binns2018and,ward2017turds}, particularly in relation to contentious issues, like the European referendum, the Brexit crisis and inequality~\cite{farrell2020vindication}. Ward and McLoughlin~\cite{ward2020turds} found previously that language that could be classified as hate-speech was rather low, however, in comparison to more generally uncivil language. Still, women from minority backgrounds were more likely to be the recipients of that type of abuse. The authors also found that men received more online abuse that was uncivil than women. Similarly to our previous work~\cite{gorrell2020politicians}, the authors demonstrated that increased name recognition and popularity had a positive relationship with levels of abuse, which may be one reason for the differences in gender. As there are more male politicians in senior roles than women, they feature more prominently and may therefore receive more abusive replies. 

Southern and Harmer \cite{southern2019twitter} conducted a deeper content analysis on tweets received by MPs during a period and found that while men received more incivility in terms of numbers of replies, women were more likely to receive an uncivil reply. Women were more likely to be stereotyped by identity (men by party) and to be questioned in their position as an MP. Gorrell \textit{et al}~\cite{gorrell2019race} noted in addition that the impacts or consequences of abusive language are not manifesting in the same ways for male and female MPs, or MPs with intersectional identities of race and gender. Where some abuse is distressing, other abuse is personal, threatening and limits women's participation in the public office~\cite{gorrell2019race,delisle2019large,pew2017}. 

Abuse toward specific parties has been difficult to distinguish, due to impacts of prominence, personal characteristics and specific events~\cite{gorrell2020politicians}. However, when controlling for this, Ward and McLoughlin~\cite{ward2020turds} found that less visible MPs had a very small percentage of hate and abuse. 
In our work, we explore some of these findings in comparison with what we can observe happening during the COVID-19 period.

\section{Data Collection and Analysis Methodology}
This study spans 1 June to 31 December 2020 inclusive, and discusses Twitter engagement with currently serving MPs that have active Twitter accounts (568 MPs in total), as well as abuse-containing replies sent to them. In total, across the seven month period, we collected and analysed 8.9 million reply tweets to the MPs, which were sent in response to the overall 545,071 tweets authored by MPs (which consist of original, retweets, and replies by MPs).   

The dataset was created by collecting tweets in real-time using Twitter's streaming API. We used the API to follow the accounts of MPs - this means we collected all the tweets sent by each MP, any replies to those tweets, and any retweets either made by the MP or of the MP's own tweets. Note that this approach does  not  collect  all  tweets  which  an  individual  would  see  in  their  timeline,  as  it does not include those in which they are just mentioned. However, ``direct replies''are included. We took this approach as the analysis results are more reliable due to the fact that replies are directed at the politician who authored the tweet, and thus, any abusive language is more likely to be directed at them. No data was lost, as volumes did not exceed Twitter rate limits at any point.

Tweets from earlier in the study have had more time to gather replies. Most replies occur in the day or two following the tweet being made, but some tweets continue to receive attention over time, and events may lead to a resurgence of focus on an earlier tweet. Reply numbers are a snapshot at the time of the study.

We analysed the dataset with the automatic abuse-based detection method developed by \cite{gorrell2020politicians}. The abuse detection method underestimates by possibly as much as a factor of two, finding more obvious verbal abuse, but missing linguistically subtler examples. This is useful for comparative findings, tracking abuse trends, and for approximation of actual abuse levels.\\

\begin{center}
\noindent\fbox{\parbox{0.95\textwidth}{
\textbf{Macro and micro averaging}\\
    In several places throughout the report, we present both a macro-average and a micro-average of abuse levels received by politicians. The micro-average is calculated on totals across all individuals. So if Corbyn receives 10 abusive tweets out of 100 and Johnson receives 15 abusive tweets out of 200, then the micro-average would be (10+15)/(100+200). The result is dominated by Johnson's counts, as he received more. \textbf{In the micro-average, a small number of individuals receiving a great many tweets may disproportionately affect the result}. In the macro-average, proportion of abuse is first calculated, and then these are averaged. So in the above example, the macro-average would be (0.1 + 0.075)/2 (because 10/100 is 0.1 and 15/200 is 0.075). \textbf{Macro-average tends to better express the experience of the average MP.}
    }
}
\end{center}

\section{Overall Abuse Levels and Main Target MPs}

\begin{center}
\noindent\fbox{\parbox{0.95\textwidth}{
    \textbf{Summary}
    \begin{itemize}
        \item Prominent politicians continue to attract the most abuse (e.g. Boris Johnson, Matt Hancock, Priti Patel, Keir Starmer), with specific events and personal characteristics or online engagement also influencing levels of abuse, as we previously reported \cite{gorrell2020politicians}
        \item The individuals receiving the most sustained abuse are Boris Johnson and Matt Hancock. In fact, Boris Johnson only fell below average abuse levels toward the end of August. Matt Hancock never falls below that average throughout the entire period studied
        \item Both Boris Johnson and Matt Hancock received their highest levels of abuse in December 2020, with the combination of Brexit negotiations and continued COVID-19 challenges intersecting. 
    \end{itemize}
    }
}
\end{center}

This section examines the overall abuse levels during the seven month period of this study and analyses reactions to members of different political parties and the UK government, as well as presents a brief gender-based comparison.

In Tables \ref{tab:overal_stat_jun_dec_2020}, \ref{tab:overal_stat_feb_may_2020}, and \ref{tab:overal_stat_nov_jun_2019}, we present an overview of our data. The columns show, for each time period, the number of original tweets authored by MPs, the number of retweets authored by them, the number of replies written by them, the number of replies received by them, number of abusive replies received by them, and abusive replies received as a percentage of all replies received by the MPs. Table \ref{tab:overal_stat_jun_dec_2020} shows the current period covered by this paper, from June - December 2020. Table \ref{tab:overal_stat_feb_may_2020} shows a comparison with the previous COVID-19 periods we studied from February - May 2020 \cite{farrell2020vindication}, Table \ref{tab:overal_stat_nov_jun_2019} shows a comparative table for periods studied before and during the 2017 and 2019 General Elections. We can see that the stress from COVID-19 and the Brexit negotiations correspond with higher levels of abuse toward British MPs during the current period studied, particularly in October and December. We can also see politicians communicating more during this period, and receiving a consistently high level of response from the public, which makes sense given the current crisis. 

\begin{table}[!htb]
\centering
\resizebox{\textwidth}{!}{%
\begin{tabular}{|l|r|r|r|r|r|r|}
\hline
\rowcolor[HTML]{EFEFEF}
\textbf{Period} &
  \textbf{\begin{tabular}[c]{@{}r@{}}Original\\ MP tweets\end{tabular}} &
  \textbf{\begin{tabular}[c]{@{}r@{}}Retweets\\ by MPs\end{tabular}} &
  \textbf{\begin{tabular}[c]{@{}r@{}}Replies \\ by MPs\end{tabular}} &
  \textbf{\begin{tabular}[c]{@{}r@{}}Replies\\ to MP\end{tabular}} &
  \textbf{\begin{tabular}[c]{@{}r@{}}Abusive Replies\\ to MPs\end{tabular}} &
  \textbf{\% Abusive} \\ \hline \hline
\textbf{Jun 2020} & 28,916           & 53,003           & 15,237          & 1,660,213          & 73,598           & 4.433                         \\ \hline
\textbf{Jul 2020} & 24,473           & 42,546           & 11,136          & 1,050,950          & 43,369           & 4.127                         \\ \hline
\textbf{Aug 2020} & 16,764           & 27,858           & 8,029           & 891,509            & 37,755           & 4.235                         \\ \hline
\textbf{Sep 2020} & 25,856           & 45,117           & 10,440          & 1,243,971          & 55,509           & 4.462                         \\ \hline
\textbf{Oct 2020} & 27,125           & 46,240           & 12,106          & 1,362,753          & 69,346           & \cellcolor[HTML]{FE7474}5.089 \\ \hline
\textbf{Nov 2020} & 27,450           & 37,965           & 11,737          & 1,348,034          & 61,421           & 4.556                         \\ \hline
\textbf{Dec 2020} & 25,159           & 35,434           & 12,480          & 1,355,797          & 73,138           & \cellcolor[HTML]{FE7474}5.394 \\ \hline \hline
\textbf{Total}    & \textbf{175,743} & \textbf{288,163} & \textbf{81,165} & \textbf{8,913,227} & \textbf{414,136} & \textbf{4.646}                \\ \hline
\end{tabular}
}
\caption{Engagement and abuse level statistics between June and Dec 2020.}
\label{tab:overal_stat_jun_dec_2020}
\end{table}

\begin{table}[!htb]
\centering
\resizebox{\textwidth}{!}{%
\begin{tabular}{|l|r|r|r|r|r|r|}
\hline
\rowcolor[HTML]{EFEFEF} 
\textbf{Period} &
  \textbf{\begin{tabular}[c]{@{}r@{}}Original\\ MP tweets\end{tabular}} &
  \textbf{\begin{tabular}[c]{@{}r@{}}Retweets\\ by MPs\end{tabular}} &
  \textbf{\begin{tabular}[c]{@{}r@{}}Replies\\ by MPs\end{tabular}} &
  \textbf{\begin{tabular}[c]{@{}r@{}}Replies\\ to MP\end{tabular}} &
  \textbf{\begin{tabular}[c]{@{}r@{}}Abusive Replies \\ to MPs\end{tabular}} &
  \textbf{\% Abusive} \\ \hline \hline
\textbf{7 Feb - 1 Mar 2020}   & 16,482 & 26,632 & 6,952  & 562,322   & 19,301 & 3.43 \\ \hline
\textbf{1 Mar - 23 Mar 2020}  & 22,419 & 39,781 & 11,482 & 777,396   & 33,069 & 4.25 \\ \hline
\textbf{23 Mar - 1 Apr}       & 11,571 & 21,821 & 7,137  & 441,983   & 13,919 & 3.15 \\ \hline
\textbf{1 Apr - 17 Apr 2020}  & 17,007 & 30,124 & 10,407 & 782,774   & 24,327 & 3.11 \\ \hline
\textbf{17 Apr - 10 May 2020} & 22,906 & 38,949 & 11,906 & 890,926   & 32,050 & 3.60 \\ \hline
\textbf{10 May - 23 May 2020} & 16,824 & 30,279 & 8,822  & 1,270,669 & 56,827 & 4.47 \\ \hline
\end{tabular}%
}
\caption{Engagement and abuse level statistics between Feb and May 2020 \cite{farrell2020vindication}.}
\label{tab:overal_stat_feb_may_2020}
\end{table}

\begin{table}[!htb]
\centering
\resizebox{\textwidth}{!}{%
\begin{tabular}{|l|r|r|r|r|r|r|}
\hline
\rowcolor[HTML]{EFEFEF} 
\textbf{Period} &
  \textbf{\begin{tabular}[c]{@{}r@{}}Original\\ MP tweets\end{tabular}} &
  \textbf{\begin{tabular}[c]{@{}r@{}}Retweets\\ by MPs\end{tabular}} &
  \textbf{\begin{tabular}[c]{@{}r@{}}Replies\\ by MPs\end{tabular}} &
  \textbf{\begin{tabular}[c]{@{}r@{}}Replies\\ to MP\end{tabular}} &
  \textbf{\begin{tabular}[c]{@{}r@{}}Abusive Replies\\ to MPs\end{tabular}} &
  \textbf{\% Abusive} \\ \hline \hline
\textbf{3 Nov - 15 Dec 2019} &   184,014 & 334,952 &  131,292 &   3,541,769 &  157,844 &   4.46 \\ \hline
\textbf{29 Apr - 9 Jun 2017} &   126,216 & 245,518 &  71,598  &   961,413   &  31,454  &   3.27 \\ \hline
\end{tabular}
}
\caption{Abuse level statistics from the 2017 and 2019 General Elections \cite{gorrell2020politicians}.}
\label{tab:overal_stat_nov_jun_2019}
\end{table}

\begin{figure}
  \includegraphics[width=.95\textwidth]{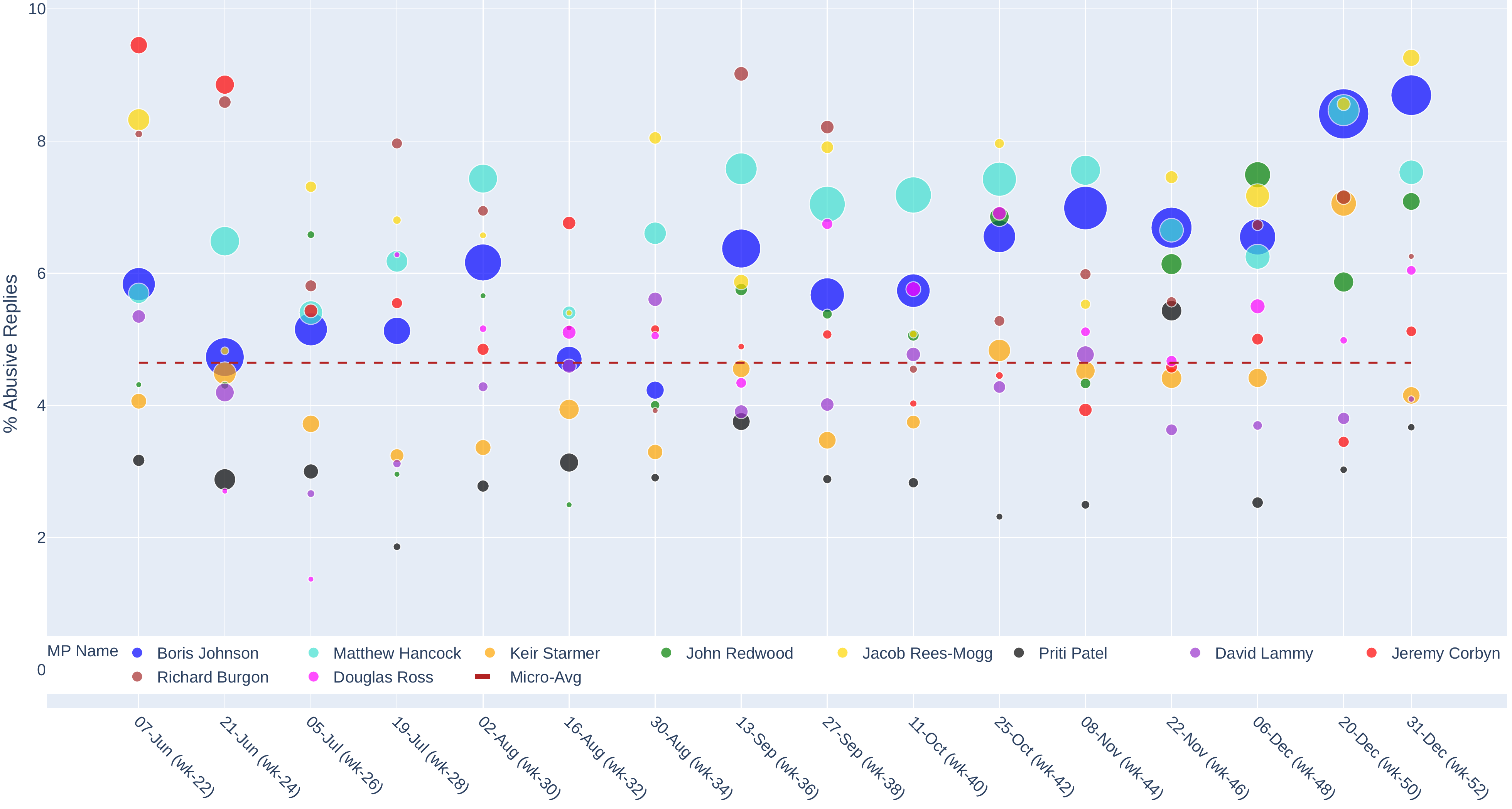}
  \caption{Top 10 most-abused MPs, from June - December 2020.}
  \label{fig:top_mps_bubble}
\end{figure}

The top 10 MPs who got the highest number of abusive replies is shown in the following bubble chart (Figure \ref{fig:top_mps_bubble}). The x-axis is the date from June to December 2020, aggregated on two-week intervals, and the y-axis corresponds to the percentage of abusive replies received, where the size of the bubble shows the absolute number of abusive replies received. We can see that, as in our previous work, those with considerable roles in the government or in the opposition parties receive many more replies, and more abusive replies than MPs with less visibility. Of the governmental figures, we see that Matt Hancock and Boris Johnson receive the most negative attention throughout, followed by Labour leader, Keir Starmer. This is to be expected as Johnson and Hancock are most visible regarding COVID-19 preparations and management, and Keir Starmer has been critical of the government response. Starmer has also received abuse from more progressive members of Labour who view Starmer as too centrist. We can expect a certain amount of party politics to play out among the Twitter users who follow any of those individuals. The last three individuals may have more specific issues impacting the levels of abuse they receive. John Redwood, an outspoken Brexiteer, came under fire for two separate issues this fall (in addition to pushback against Brexit). First, he received rebukes for suggesting that investors take their money outside of the UK \footnote{\url{https://www.forbes.com/sites/francescoppola/2017/11/12/british-lawmaker-advises-investors-to-take-their-money-out-of-the-uk/?sh=33947d9a4c1e}}. Then, after a report was released detailing the ways in which several conservative MPs (including Redwood) have profited from privatization in the NHS and from the COVID-19 crisis \footnote{\url{https://www.thelondoneconomic.com/opinion/revealed-the-links-between-tory-mps-and-the-people-profiting-from-nhs-privatisation-213827}}, Redwood received considerable criticism. Jacob Rees-Mogg, another figure who is polarising in the British public, was also implicated in COVID-related profiteering. However, these peaks may be explained by Rees-Mogg's campaign and subsequent speech in parliament last June on returning MPs to the chamber\footnote{\url{https://www.gov.uk/government/speeches/leader-of-the-house-of-commons-speech-8-june-2020}}. What became known as the ``Mogg-Conga''\footnote{\url{https://www.bbc.co.uk/news/uk-politics-52897865}} (also in some of our hashtag analysis), refers to the way that members were required to file  into the building to vote, following social distancing guidelines. Rees-Mogg was also involved in a public critique of Unicef, which has offered to provide free meals to school children, when it appeared the British parliament would not provide them. Mogg accused Unicef of ``playing politics''\footnote{\url{https://www.bbc.co.uk/news/uk-55354958}}. Priti Patel has typically attracted abuse for strong language around migration policies. In the previous COVID-19 periods, Patel was accused of bullying, a charge which has followed her into the current period, after Boris Johnson chose to keep her in her role\footnote{\url{https://www.bbc.co.uk/news/uk-55026137}}.

\begin{figure}
  \includegraphics[width=.95\textwidth]{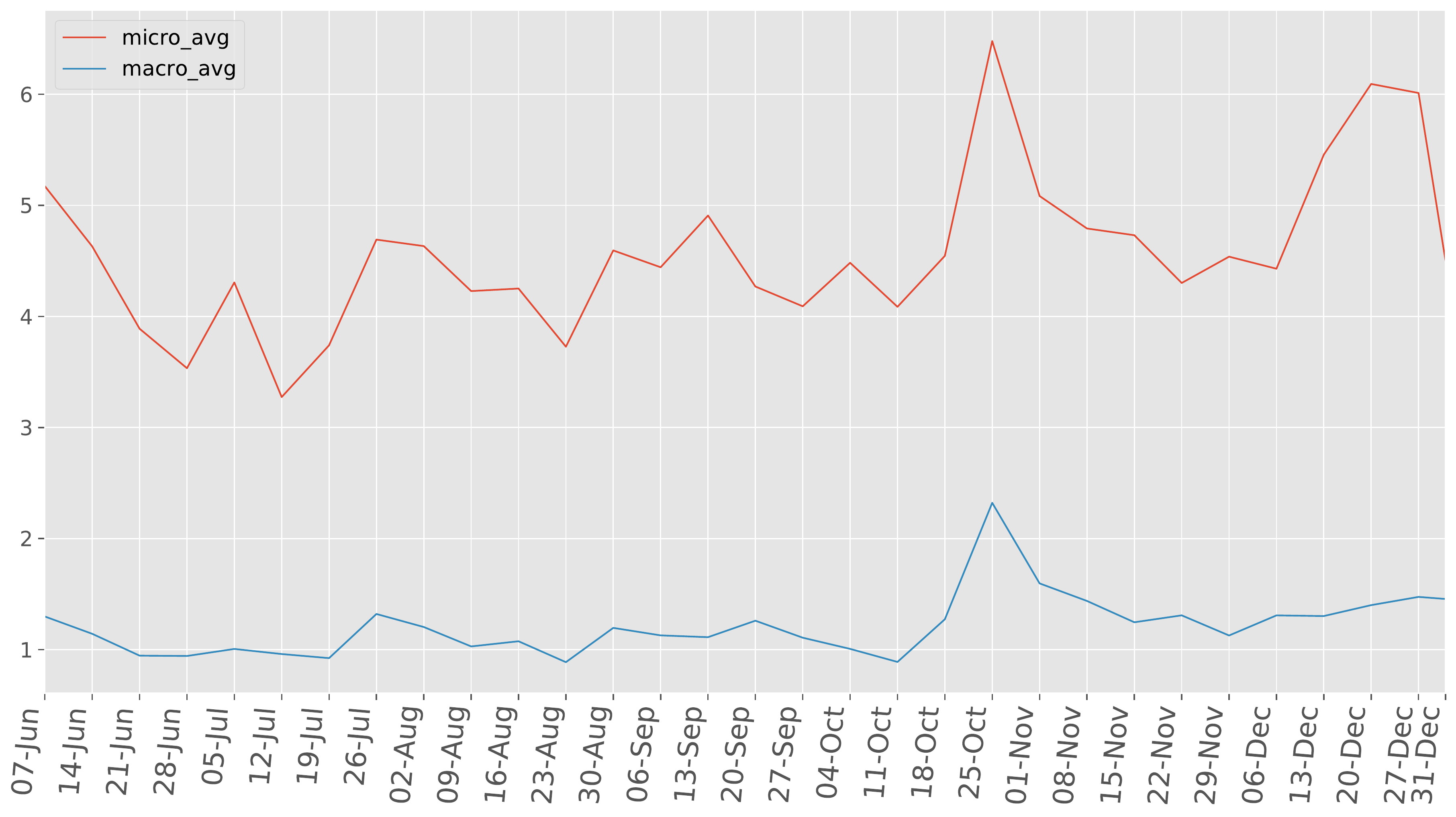}
  \caption{All MPs, macro and micro abuse levels, per week.}
  \label{fig:all_macro_micro}
\end{figure}

The timeline in Figure \ref{fig:all_macro_micro} shows abuse levels overall, toward all MPs. It reflects a per-week basis since the beginning of June. From the beginning of June it shows a slight decrease through the rest of June, with a smaller peak at the beginning of July, possibly as UK businesses cut more than 11,000 jobs in 48 hours. However, by "Super Saturday" on July 4th, when pubs and restaurants are able to reopen, abuse levels dip back under 4\%. This effect is short-lived, as abuse levels rise again through the second half of July. Compulsory mask wearing was slowly introduced during this time. However, this took place at different paces across the four nations, which may have led to some confusion. There are likely to be regional differences as well, as several lockdowns were introduced in the northern part of England. Easing of restrictions was also postponed in some areas. 

Abuse levels remain more or less steady through the rest of the summer and fall with a sharp rise from the 18th - 25th of October. During this time, PM Boris Johnson was in a public dispute with Manchester mayor Andy Burnham over financial support during the lockdown. London, as well, was put under increased restriction. Across the four nations the difference in response was quite stark. Scotland introduced a 5 tier system. Wales had the firebreak lockdown. Anger over lockdown boiled over into protests on the 24th in London with tens of thousands of participants. In addition, rows over the government's decision not to  extend free school meals to children in England continued over the month. However, upon further analysis of the tweets, as our analysis above can confirm, we linked the peak of abuse in October to another incident in which Angela Rayner referred to Chris Clarkson as `scum', while he was speaking in parliament on the 21st of October. A few hours later, Amanda Milling tweeted that this was unacceptable behaviour\footnote{\url{https://twitter.com/amandamilling/status/1318920363473047552}}. This tweet got a number of abusive replies. Then, on the 23rd, Ms. Milling tweeted a request for the Labour party to ``take action against Labour MPs and party members who perpetrate abuse'', which resulted in even more abuse. Chris Clarkson tweeted his appreciation for her support\footnote{\url{https://twitter.com/ChrisClarksonMP/status/1318953597959393280}}, which also received a number of abusive replies.  Interestingly, when Angela Rayner tweeted on 21 Oct at 18:45, the amount of abuse she received was relatively low (219 of 1550) in comparison to Amanda Milling's tweet, given that Rayner had already abused Chris Clarkson by that point.

Abuse then remains elevated at between 4\% and 5\% across November with another sharp rise across December, which peaks around the 20th and is maintained across the holiday period. With the Brexit deadline coming quickly into focus, by mid-month, 68\% of the country was on the toughest restrictions and yet the government was still promising an easing of restrictions over the holidays. Then, the new strain in the UK was discovered mid-month, and the introduction of Tier 4 restrictions on the 19th, `canceled' holiday plans for many in England. The three other nations made similar changes to their holiday restrictions. While abuse levels appeared to be falling by the very end of the month, potentially as Brexit negotiations were clarified, the effects of January's lockdown (which effectively has kept many in the UK on a persistent lockdown since late October) will be interesting to observe in future work.

\section{Long-term Context}

\begin{figure}
  \includegraphics[width=.95\textwidth]{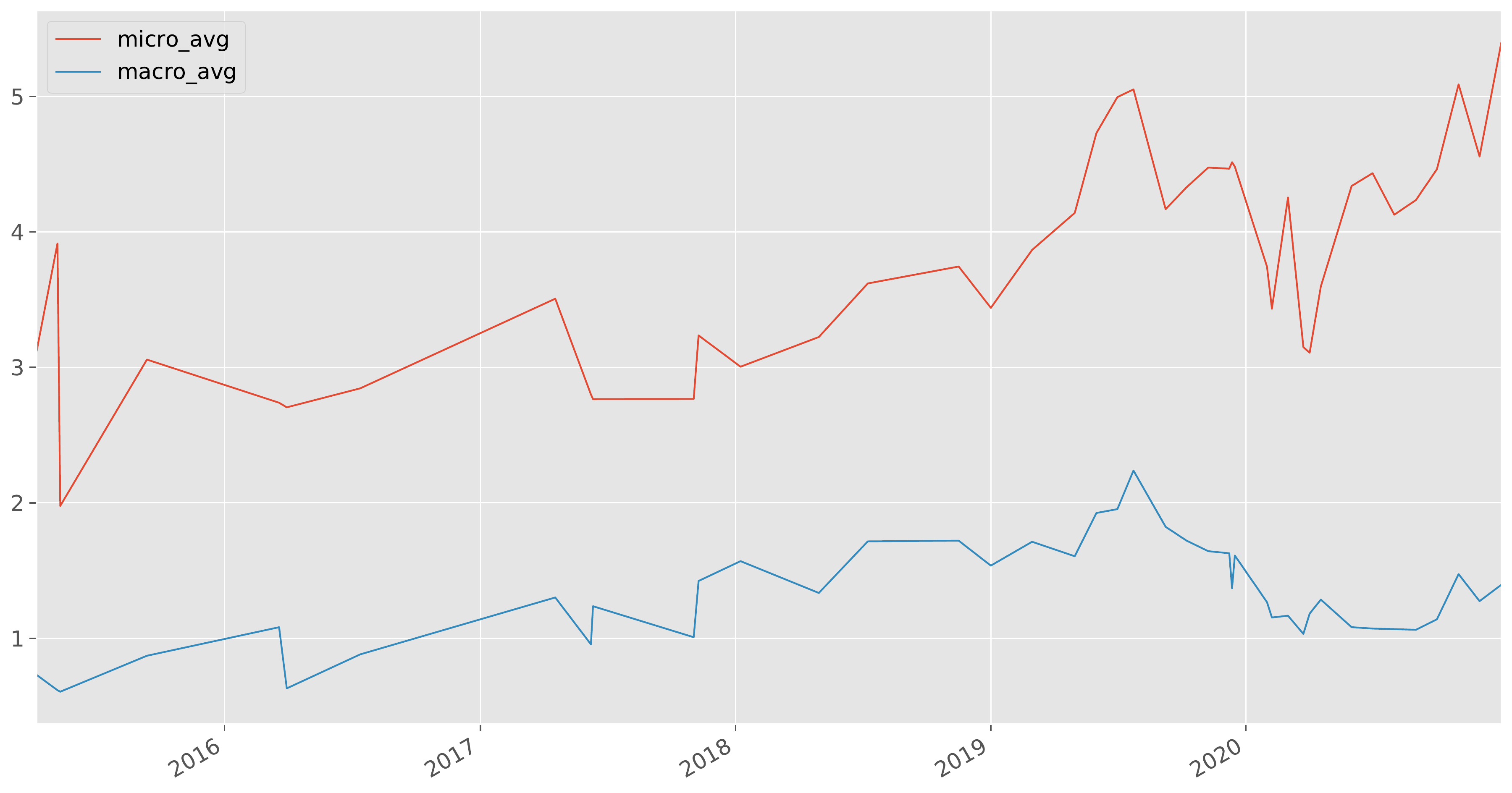}
  \caption{Timeline of abuse received by MPs from 2015 until end of 2020.}
  \label{fig:tl2015_2020}
\end{figure}

To understand the level of abuse received by MPs during the COVID-19 crisis, it is helpful to make a comparison across all of the time periods we have studied, from General Elections in 2015, 2017 and 2019, all the way through the COVID-19 periods of our previous and current work. From the timeline shown in Figure \ref{fig:tl2015_2020}, we see that aside from a blip around the 2015 general election, abuse toward MPs on Twitter has been tending to rise from a minimum of 2\% of replies in 2015, peaking mid-2019 at over 5\%  with a smaller peak of around 4.5\% around the 2019 general election. We can see that toward the end of 2020, abuse levels had reached their highest point yet, at 5.4\%.

\section{Top Recipients of Abuse per Party and Gender}
In Table \ref{tab:toprecipientsOfAbuse}, we see the top 10 MPs receiving abusive replies each month of the studied period from June - December 2020. The numbers in brackets are showing first the number of abusive replies against the number of all replies received. So, for example, Boris Johnson received 12,768 abusive replies in June, out of 252,271 replies in total. 
\begin{table}[!htb]
\centering
\resizebox{\textwidth}{!}{%
\begin{tabular}{|lllllll|}
\rowcolor[HTML]{EFEFEF} 
\hline
\textbf{June} &
  \textbf{July} &
  \textbf{August} &
  \textbf{September} &
  \textbf{October} &
  \textbf{November} &
  \textbf{December} \\ \hline\hline
\rowcolor[HTML]{CFE2F3} 
Boris Johnson &
  Boris Johnson &
  Boris Johnson &
  Boris Johnson &
  Matthew Hancock &
  Boris Johnson &
  Boris Johnson \\
\rowcolor[HTML]{CFE2F3} 
(12786/252271) &
  (9487/168847) &
  (4837/99519) &
  (12200/196481) &
  (10146/137162) &
  (14880/218905) &
  (18145/223108) \\
\rowcolor[HTML]{CFE2F3} 
Matthew Hancock &
  Matthew Hancock &
  Matthew Hancock &
  Matthew Hancock &
  Boris Johnson &
  Matthew Hancock &
  Matthew Hancock \\
\rowcolor[HTML]{CFE2F3} 
(6636/111073) &
  (5796/85779) &
  (2851/45756) &
  (10405/142332) &
  (9987/158993) &
  (5922/85294) &
  (6951/92418) \\
\rowcolor[HTML]{F4CCCC} 
Keir Starmer &
  \cellcolor[HTML]{FCE5CD}Ed Davey &
  Keir Starmer &
  Keir Starmer &
  Keir Starmer &
  \cellcolor[HTML]{CFE2F3}John Redwood &
  Keir Starmer \\
\rowcolor[HTML]{F4CCCC} 
(3986/94082) &
  \cellcolor[HTML]{FCE5CD}(2387/19784) &
  (2719/75431) &
  (2518/64320) &
  (3606/77218) &
  \cellcolor[HTML]{CFE2F3}(3876/59347) &
  (4742/83357) \\
\rowcolor[HTML]{F4CCCC} 
Barry Gardiner &
  Keir Starmer &
  Dawn Butler &
  \cellcolor[HTML]{CFE2F3}Jacob Rees-Mogg &
  \cellcolor[HTML]{CFE2F3}Amanda Milling &
  Keir Starmer &
  \cellcolor[HTML]{CFE2F3}John Redwood \\
\rowcolor[HTML]{F4CCCC} 
(3213/17272) &
  (1962/58901) &
  (2556/58119) &
  \cellcolor[HTML]{CFE2F3}(1736/27113) &
  \cellcolor[HTML]{CFE2F3}(2854/19263) &
  (2883/71112) &
  \cellcolor[HTML]{CFE2F3}(3623/56227) \\
\rowcolor[HTML]{CFE2F3} 
Priti Patel &
  \cellcolor[HTML]{F4CCCC}Barry Sheerman &
  Iain Duncan Smith &
  \cellcolor[HTML]{F4CCCC}Richard Burgon &
  John Redwood &
  Priti Patel &
  Jacob Rees-Mogg \\
\rowcolor[HTML]{CFE2F3} 
(3015/104071) &
  \cellcolor[HTML]{F4CCCC}(1779/21412) &
  (2145/24143) &
  \cellcolor[HTML]{F4CCCC}(1680/20029) &
  (2330/37491) &
  (2008/47080) &
  (3200/41901) \\
\rowcolor[HTML]{CFE2F3} 
\cellcolor[HTML]{F4CCCC}Jeremy Corbyn &
  Dominic Raab &
  \cellcolor[HTML]{F4CCCC}David Lammy &
  Priti Patel &
  Chris Clarkson &
  Jacob Rees-Mogg &
  Imran Ahmad-Khan \\
\rowcolor[HTML]{CFE2F3} 
\cellcolor[HTML]{F4CCCC}(2957/35489) &
  (1557/31631) &
  \cellcolor[HTML]{F4CCCC}(1773/36299) &
  (1653/46795) &
  (2036/6762) &
  (1989/26300) &
  (2605/25118) \\
\rowcolor[HTML]{F4CCCC} 
David Lammy &
  Jeremy Corbyn &
  \cellcolor[HTML]{CFE2F3}Priti Patel &
  David Lammy &
  \cellcolor[HTML]{CFE2F3}Douglas Ross &
  David Lammy &
  \cellcolor[HTML]{CFE2F3}Grant Shapps \\
\rowcolor[HTML]{F4CCCC} 
(2314/54385) &
  (1527/28144) &
  \cellcolor[HTML]{CFE2F3}(1766/58021) &
  (1598/39304) &
  \cellcolor[HTML]{CFE2F3}(1701/27253) &
  (1686/40485) &
  \cellcolor[HTML]{CFE2F3}(1427/24171) \\
\rowcolor[HTML]{CFE2F3} 
Jacob Rees-Mogg &
  Priti Patel &
  Gavin Williamson &
  Andrea Jenkyns &
  Selaine Saxby &
  Douglas Ross &
  Andrew Bridgen \\
\rowcolor[HTML]{CFE2F3} 
(2293/30434) &
  (1199/43613) &
  (1338/20479) &
  (1227/22194) &
  (1675/17073) &
  (1314/26717) &
  (1394/13052) \\
\rowcolor[HTML]{CFE2F3} 
\cellcolor[HTML]{F4CCCC}Dawn Butler &
  \cellcolor[HTML]{F4CCCC}Richard Burgon &
  Douglas Ross &
  John Redwood &
  Rishi Sunak &
  Andrew Rosindell &
  Michael Fabricant \\
\rowcolor[HTML]{CFE2F3} 
\cellcolor[HTML]{F4CCCC}(1680/26013) &
  \cellcolor[HTML]{F4CCCC}(1066/14754) &
  (1288/26197) &
  (1052/18383) &
  (1648/35925) &
  (1137/8390) &
  (1356/23872) \\
\rowcolor[HTML]{CFE2F3} 
Dominic Raab &
  Jacob Rees-Mogg &
  \cellcolor[HTML]{F4CCCC}Zarah Sultana &
  Kevin Hollinrake &
  \cellcolor[HTML]{F4CCCC}David Lammy &
  Dominic Raab &
  \cellcolor[HTML]{F4CCCC}Jeremy Corbyn \\
\rowcolor[HTML]{CFE2F3} 
(1647/30878) &
  (905/12311) &
  \cellcolor[HTML]{F4CCCC}(1193/24739) &
  (1004/7669) &
  \cellcolor[HTML]{F4CCCC}(1491/32903) &
  (1136/19473) &
  \cellcolor[HTML]{F4CCCC}(1288/28999) \\
\rowcolor[HTML]{CFE2F3} 
\cellcolor[HTML]{FCE5CD}Layla Moran &
  Rishi Sunak &
  \cellcolor[HTML]{F4CCCC}Neil Coyle &
  Douglas Ross &
  Nadine Dorries &
  Iain Duncan Smith &
  \cellcolor[HTML]{F4CCCC}Richard Burgon \\
\rowcolor[HTML]{CFE2F3} 
\cellcolor[HTML]{FCE5CD}(1565/22686) &
  (783/34638) &
  \cellcolor[HTML]{F4CCCC}(1115/10456) &
  (943/16860) &
  (1342/33851) &
  (1032/20303) &
  \cellcolor[HTML]{F4CCCC}(1141/15966) \\
\rowcolor[HTML]{F4CCCC} 
Nadia Whittome &
  \cellcolor[HTML]{CFE2F3}James Cleverly &
  Jeremy Corbyn &
  Dawn Butler &
  Angela Rayner &
  Jeremy Corbyn &
  David Lammy \\
\rowcolor[HTML]{F4CCCC} 
(1548/23351) &
  \cellcolor[HTML]{CFE2F3}(727/10858) &
  (1096/17997) &
  (906/19957) &
  (1340/25246) &
  (994/22586) &
  (1049/26042) \\
\rowcolor[HTML]{CFE2F3} 
Therese Coffey &
  \cellcolor[HTML]{F4CCCC}David Lammy &
  Michael Gove &
  Dehenna Davison &
  Johnny Mercer &
  \cellcolor[HTML]{FFF2CC}Ian Blackford &
  Tobias Ellwood \\
\rowcolor[HTML]{CFE2F3} 
(1451/22069) &
  \cellcolor[HTML]{F4CCCC}(707/19524) &
  (827/10663) &
  (798/19249) &
  (1287/13878) &
  \cellcolor[HTML]{FFF2CC}(943/14477) &
  (1007/13140) \\
\rowcolor[HTML]{CFE2F3} 
Sajid Javid &
  Andrea Jenkyns &
  Jacob Rees-Mogg &
  Lucy Allan &
  \cellcolor[HTML]{FFF2CC}Margaret Ferrier &
  \cellcolor[HTML]{F4CCCC}Richard Burgon &
  Lucy Allan \\
\rowcolor[HTML]{CFE2F3} 
(1197/24873) &
  (450/7156) &
  (736/9804) &
  (709/19163) &
  \cellcolor[HTML]{FFF2CC}(1185/16501) &
  \cellcolor[HTML]{F4CCCC}(915/15125) &
  (994/16564) \\
\rowcolor[HTML]{CFE2F3} 
\cellcolor[HTML]{F4CCCC}Richard Burgon &
  Liz Truss &
  James Cleverly &
  Nadine Dorries &
  James Cleverly &
  Rishi Sunak &
  \cellcolor[HTML]{FCE5CD}Ed Davey \\
\rowcolor[HTML]{CFE2F3} 
\cellcolor[HTML]{F4CCCC}(1159/16341) &
  (440/16956) &
  (544/9580) &
  (603/13568) &
  (1175/17902) &
  (895/33440) &
  \cellcolor[HTML]{FCE5CD}(964/15233) \\
\rowcolor[HTML]{CFE2F3} 
Michael Gove &
  Caroline Nokes &
  \cellcolor[HTML]{F4CCCC}Nadia Whittome &
  \cellcolor[HTML]{F4CCCC}Jeremy Corbyn &
  Sajid Javid &
  Nadine Dorries &
  Douglas Ross \\
\rowcolor[HTML]{CFE2F3} 
(1080/14429) &
  (437/9181) &
  \cellcolor[HTML]{F4CCCC}(478/5813) &
  \cellcolor[HTML]{F4CCCC}(571/11299) &
  (1066/17716) &
  (858/18395) &
  (893/15740) \\
\rowcolor[HTML]{CFE2F3} 
\cellcolor[HTML]{F4CCCC}Jess Phillips &
  Esther McVey &
  Natalie Elphicke &
  Scott Benton &
  \cellcolor[HTML]{F4CCCC}Richard Burgon &
  Michael Gove &
  David Davis \\
\rowcolor[HTML]{CFE2F3} 
\cellcolor[HTML]{F4CCCC}(1061/28876) &
  (404/5459) &
  (471/8285) &
  (569/6269) &
  \cellcolor[HTML]{F4CCCC}(936/18529) &
  (810/10921) &
  (829/10318) \\
\rowcolor[HTML]{CFE2F3} 
Henry Smith &
  Nadine Dorries &
  John Redwood &
  Johnny Mercer &
  \cellcolor[HTML]{F4CCCC}Jeremy Corbyn &
  Scott Benton &
  Priti Patel \\
\rowcolor[HTML]{CFE2F3} 
(913/6372) &
  (394/11290) &
  (465/13131) &
  (544/7364) &
  \cellcolor[HTML]{F4CCCC}(928/23536) &
  (804/7511) &
  (806/25167) \\
\cellcolor[HTML]{F4CCCC}Naz Shah &
  \cellcolor[HTML]{CFE2F3}Jeremy Hunt &
  \cellcolor[HTML]{FCE5CD}Layla Moran &
  \cellcolor[HTML]{FFF2CC}Ian Blackford &
  \cellcolor[HTML]{CFE2F3}Steve Baker &
  \cellcolor[HTML]{F4CCCC}Barry Sheerman &
  \cellcolor[HTML]{FFF2CC}Ian Blackford \\
\cellcolor[HTML]{F4CCCC}(843/10900) &
  \cellcolor[HTML]{CFE2F3}(391/7877) &
  \cellcolor[HTML]{FCE5CD}(419/7664) &
  \cellcolor[HTML]{FFF2CC}(528/13009) &
  \cellcolor[HTML]{CFE2F3}(916/22494) &
  \cellcolor[HTML]{F4CCCC}(766/8392) &
  \cellcolor[HTML]{FFF2CC}(784/13293) \\
\rowcolor[HTML]{CFE2F3} 
Nadine Dorries &
  \cellcolor[HTML]{F4CCCC}Jess Phillips &
  Jeremy Hunt &
  Tobias Ellwood &
  Priti Patel &
  Joy Morrissey &
  Gavin Williamson \\
\rowcolor[HTML]{CFE2F3} 
(793/20349) &
  \cellcolor[HTML]{F4CCCC}(377/14320) &
  (376/4509) &
  (526/11134) &
  (763/28733) &
  (638/13718) &
  (683/10389) \\ \hline
\multicolumn{7}{c}{\textbf{Colour Codes}} \\
\cellcolor[HTML]{F4CCCC}\textbf{Labour} &
  \cellcolor[HTML]{CFE2F3}\textbf{Conservatives} &
  \cellcolor[HTML]{FCE5CD}\textbf{LibDems} &
  \cellcolor[HTML]{D9EAD3}\textbf{Green} &
  \cellcolor[HTML]{FFF2CC}\textbf{SNP} &
  \cellcolor[HTML]{D9D9D9}\textbf{DUP} &
  \cellcolor[HTML]{D9D2E9}\textbf{Plaid Cymru}
\end{tabular}%
}
\caption{Top 10 MPs receiving the most abusive replies from June - December 2020}
\label{tab:toprecipientsOfAbuse}
\end{table}

As we can see from the table, the Conservatives received the most significant numbers of abusive replies in this period, with the top two spots occupied by Boris Johnson and Matt Hancock. We can see the impact of the leadership contest in the Liberal Democrats, as well as a few outspoken members of the SNP, but otherwise, the smaller parties did not feature on this list. This doesn't mean that they do not receive abusive replies. In contrast with the bigger parties, however, it is less noticeable in the larger trends and patterns we can observe. In the next two sections, we discuss the influence of gender and party affiliation on receiving abusive replies during COVID-19. 

\subsection{Difference in responses to different parties}
We can see from tables \ref{tab:topAbusedConservative} - \ref{tab:topAbusedLibDems} that the Tory party received the highest percentage of abusive replies from July 2020 onwards, which stays above 5\% starting from September 2020 onwards, as the COVID-19 crisis deepened and the Brexit negotiations with the EU started nearing completion. In contrast, the percentage of abuse received by Labour MPs remained below 4\% July 2020 onwards, continuing the trend observed from April 2020. As we argued in our previous work \cite{farrell2020vindication}, the attention on the Tory party most likely has to do with a combination of the conservatives being in power during a significant crisis and the general uncertainty in current events, with which the public is largely uncomfortable. However, in this period, we also have the first reports of the consequences of the pandemic on the job market \cite{Mayhew2020}, the economy \cite{David2020}, household income \cite{Brewer2020}, and mental health \cite{Johnson2021, White2020}, for example, which may be influencing public perception of how the Tories have managed the crisis. The Liberal Democrats have a spike in abuse in July, most likely reflecting confusion around the leadership contest, which was first postponed to May 2021\footnote{\url{https://www.libdems.org.uk/leadership-election-postponed}}. After a number of complaints from party members, this decision was reversed and the election proceeded through July and August 2020 \footnote{\url{https://www.libdems.org.uk/leadership-timetable}}. Though the smaller parties do not receive a large portion of abusive replies, in August, we saw a surge of abuse toward the Democratic Unionist Party, potentially toward Sammy Wilson, who was in conflict with the government over Brexit in August 2020. As the Brexit crisis comes to an end, abuse levels appear to level out alongside the SNP.  

\begin{table}[!htb]
\centering
\resizebox{0.75\textwidth}{!}{%
\begin{tabular}{|lrrrr|}
\hline
\rowcolor[HTML]{EFEFEF} 
\textbf{MP Name} &
  \textbf{\begin{tabular}[c]{@{}r@{}}Original \\ MP tweets\end{tabular}} &
  \textbf{\begin{tabular}[c]{@{}r@{}}Replies \\ to MP\end{tabular}} &
  \textbf{\begin{tabular}[c]{@{}r@{}}Abusive Replies \\ to MPs\end{tabular}} &
  \textbf{\% Abusive} \\ \hline \hline
Boris Johnson     & 395   & 1,318,124 & 82,322 & 6.245  \\ 
Matthew Hancock   & 702   & 699,814   & 48,707 & 6.960  \\ 
John Redwood      & 291   & 199,893   & 12,105 & 6.056  \\ 
Jacob Rees-Mogg   & 159   & 155,656   & 11,396 & 7.321  \\ 
Priti Patel       & 170   & 353,480   & 11,210 & 3.171  \\ 
Douglas Ross      & 441   & 115,775   & 6,319  & 5.458  \\ 
Dominic Raab      & 455   & 120,614   & 5,568  & 4.616  \\ 
Rishi Sunak       & 264   & 171,044   & 4,894  & 2.861  \\ 
Iain Duncan Smith & 408   & 79,741    & 4,867  & 6.104  \\ 
Nadine Dorries    & 324   & 117,348   & 4,854  & 4.136  \\ 
James Cleverly    & 420   & 83,338    & 4,360  & 5.232  \\ 
Michael Gove      & 56    & 59,134    & 4,018  & 6.795  \\ 
Andrea Jenkyns    & 210   & 64,860    & 3,797  & 5.854  \\ 
Amanda Milling    & 341   & 29,175    & 3,137  & 10.752 \\ 
Steve Baker       & 1,059 & 98,675    & 2,959  & 2.999  \\ 
Sajid Javid       & 315   & 58,193    & 2,806  & 4.822  \\ 
Tobias Ellwood    & 254   & 53,636    & 2,804  & 5.228  \\ 
Imran Ahmad-Khan  & 245   & 29,996    & 2,786  & 9.288  \\ 
Gavin Williamson  & 71    & 46,692    & 2,620  & 5.611  \\ 
Andrew Bridgen    & 132   & 29,885    & 2,617  & 8.757  \\ \hline 
\end{tabular}%
}
\caption{Conservative MPs who had the highest percentage of abusive replies from June to December 2020.}
\label{tab:topAbusedConservative}
\end{table}

\begin{table}[!htb]
\centering
\resizebox{0.75\textwidth}{!}{%
\begin{tabular}{|lrrrr|}
\hline
\rowcolor[HTML]{EFEFEF} 
\textbf{MP Name} &
  \textbf{\begin{tabular}[c]{@{}r@{}}Original \\ MP tweets\end{tabular}} &
  \textbf{\begin{tabular}[c]{@{}r@{}}Replies \\ to MP\end{tabular}} &
  \textbf{\begin{tabular}[c]{@{}r@{}}Abusive Replies \\ to MPs\end{tabular}} &
  \textbf{\% Abusive} \\ \hline \hline
Keir Starmer      & 572   & 524,421 & 22,416 & 4.274  \\
David Lammy       & 775   & 248,942 & 10,618 & 4.265  \\
Jeremy Corbyn     & 385   & 168,050 & 9,361  & 5.570  \\
Richard Burgon    & 646   & 104,105 & 7,138  & 6.857  \\
Dawn Butler       & 726   & 130,644 & 5,985  & 4.581  \\
Barry Sheerman    & 2,564 & 60,498  & 4,220  & 6.975  \\
Zarah Sultana     & 621   & 103,519 & 3,869  & 3.737  \\
Angela Rayner     & 1,353 & 120,060 & 3,420  & 2.849  \\
Barry Gardiner    & 109   & 18,664  & 3,257  & 17.451 \\
Nadia Whittome    & 452   & 56,266  & 2,984  & 5.303  \\
Jess Phillips     & 742   & 111,724 & 2,950  & 2.640  \\
Diane Abbott      & 411   & 56,470  & 1,751  & 3.101  \\
Neil Coyle        & 1,188 & 21,220  & 1,620  & 7.634  \\
Lisa Nandy        & 363   & 59,803  & 1,566  & 2.619  \\
Naz Shah          & 220   & 19,115  & 1,248  & 6.529  \\
Chris Bryant      & 1,416 & 41,493  & 1,161  & 2.798  \\
Wes Streeting     & 1,153 & 32,455  & 1,043  & 3.214  \\
John McDonnell    & 408   & 34,375  & 1,018  & 2.961  \\
Jon Ashworth      & 472   & 28,676  & 946    & 3.299  \\
Rosena Allin-Khan & 676   & 43,276  & 836    & 1.932 \\ \hline
\end{tabular}%
}
\caption{Labour MPs who had the highest percentage of abusive replies from June to December 2020.}
\label{tab:topAbusedLabour}
\end{table}

\begin{table}[!htb]
\centering
\resizebox{0.75\textwidth}{!}{%
\begin{tabular}{|lrrrr|}
\hline
\rowcolor[HTML]{EFEFEF} 
\textbf{MP Name} &
  \textbf{\begin{tabular}[c]{@{}r@{}}Original \\ MP tweets\end{tabular}} &
  \textbf{\begin{tabular}[c]{@{}r@{}}Replies \\ to MP\end{tabular}} &
  \textbf{\begin{tabular}[c]{@{}r@{}}Abusive Replies \\ to MPs\end{tabular}} &
  \textbf{\% Abusive} \\ \hline \hline
Ed Davey            & 758   & 63,686 & 4,942 & 7.760 \\
Layla Moran         & 1,221 & 42,654 & 2,425 & 5.685 \\
Wera Hobhouse       & 569   & 9,151  & 342   & 3.737 \\
Tim Farron          & 623   & 14,299 & 294   & 2.056 \\
Munira Wilson       & 607   & 9,425  & 152   & 1.613 \\
Daisy Cooper        & 389   & 6,955  & 151   & 2.171 \\
Christine Jardine   & 571   & 3,983  & 67    & 1.682 \\
Sarah Olney         & 235   & 3,397  & 66    & 1.943 \\
Jamie Stone         & 539   & 7,813  & 29    & 0.371 \\
Alistair Carmichael & 187   & 1,960  & 19    & 0.969 \\
Wendy Chamberlain   & 237   & 906    & 3     & 0.331\\ \hline
\end{tabular}%
}
\caption{Liberal Democrats' MPs who had the highest percentage of abusive replies from June to December 2020.}
\label{tab:topAbusedLibDems}
\end{table}

\subsection{Differences in abuse based on gender}
\begin{figure}
  \includegraphics[width=.85\textwidth]{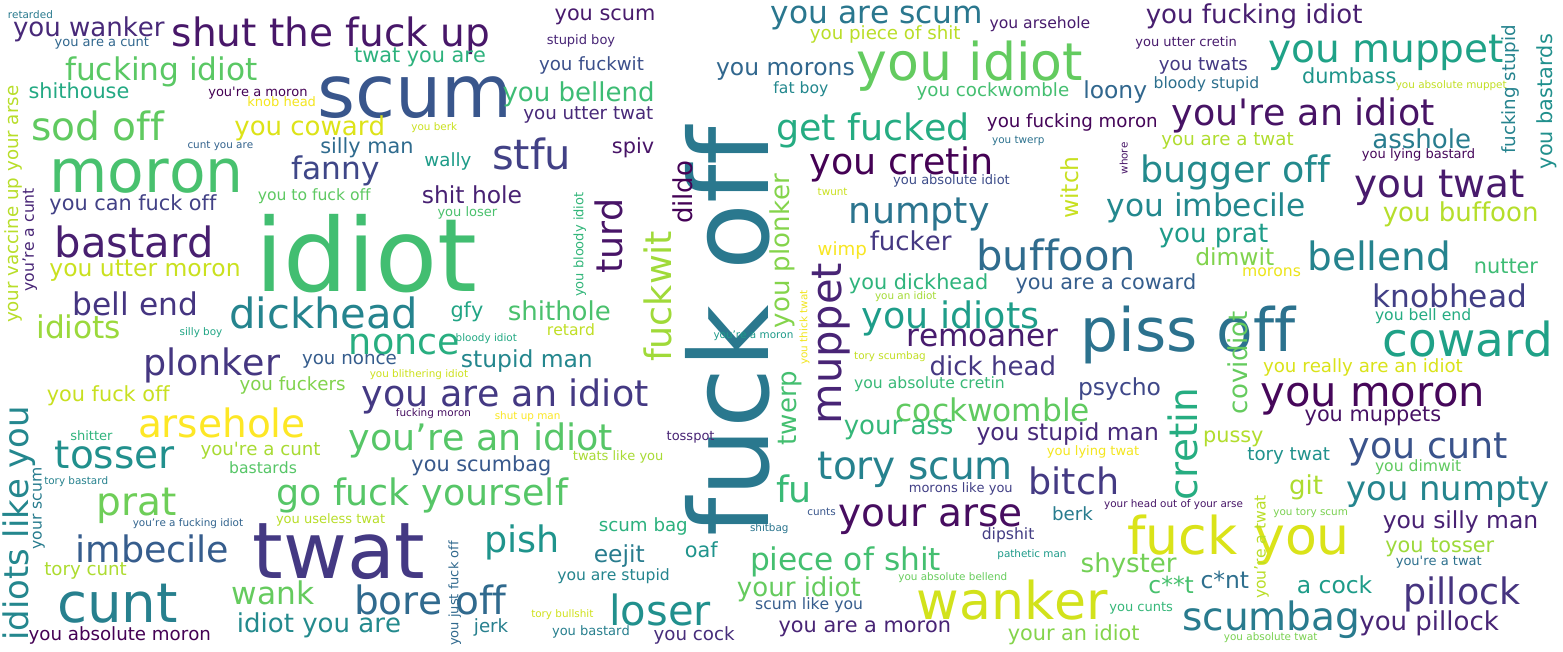}
  \caption{Word cloud of abuse terms towards male MPs.}
  \label{fig:wc_male_abuse}
\end{figure}

\begin{figure}
  \includegraphics[width=.85\textwidth]{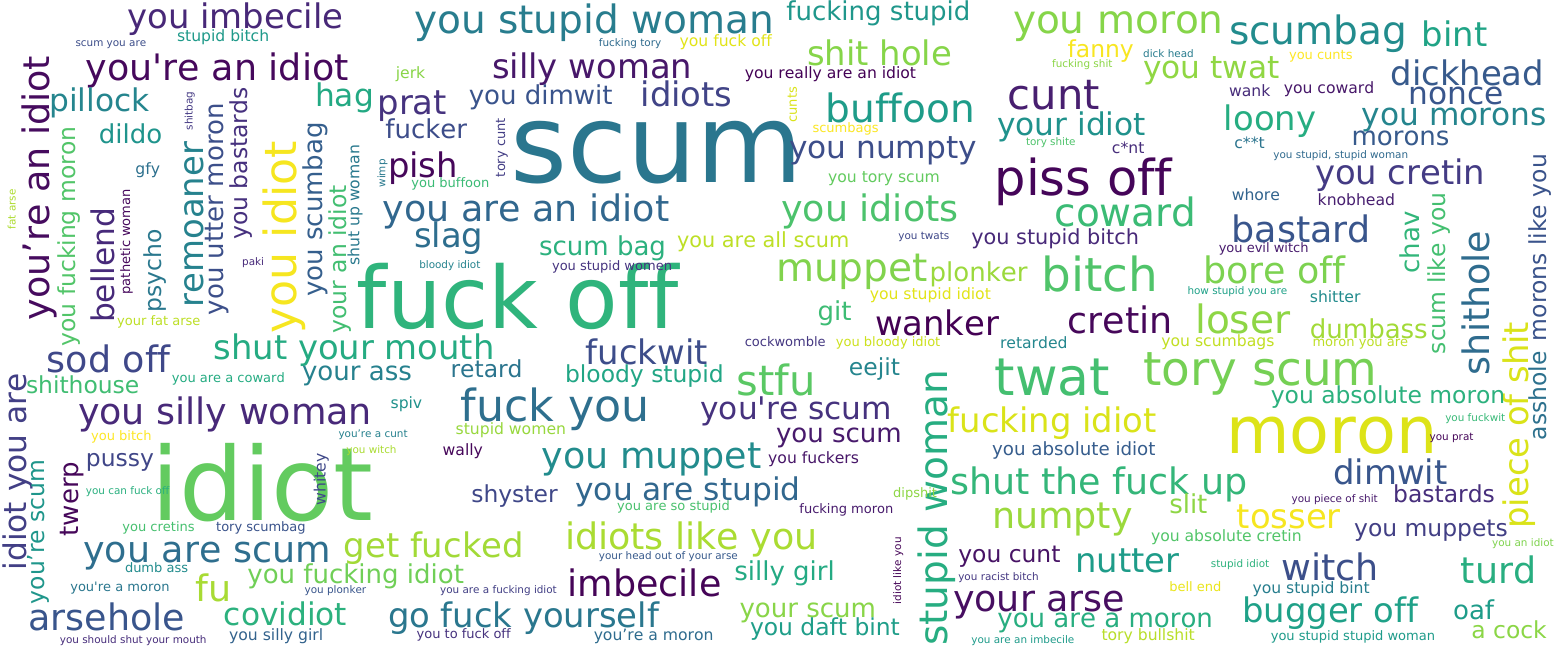}
  \caption{Word cloud of abuse terms towards female MPs.}
  \label{fig:wc_female_abuse}
\end{figure}

Violence against women in politics is an established issue. A 2016 study indicated that a quarter of women politicians had received some type of physical violence, and a fifth some time of sexual violence, globally \cite{akhtar2019prevalence}. Studying instances of online violence against women are intended to investigate this specific issue we see playing out in the physical world, to see if we can identify additional features of suppression or exclusion of women from politics in the online space. However, large-scale analysis of online hate and abusive language in the UK have not returned significant differences for men and women \cite{vidgen2019trajectories}. Indeed, at first glance, our analysis of abusive terms directed at male (Figure \ref{fig:wc_male_abuse}) and female MPs (Figure \ref{fig:wc_female_abuse}) appear to confirm this. The reasons for this could be diverse. Lexical approaches may not capture subtler forms of discrimination \cite{gorrell2020politicians}. Other features may play an important role from an intersectional perspective. For example, our work indicated that prominence and personal characteristics are important features in online abuse. Studies from similar contexts have suggested that gender may play a more prominent role when a woman is a very visible government figure \cite{rheault2019politicians}. 

We know the gender identity of MPs in the UK through self-report or use of pronouns in the media. All MPs fall into binary gender classification at the moment. We can see from our Table \ref{tab:topSexistAbuse} in the analysis that, when we looked for sexist language that focuses on the gender of any MP, only four men feature on our top list despite having a much higher representation in the UK political context. The women on the list come from all major parties and the SNP. Some are quite visible on Twitter, as seen in the number of tweets they sent during the time, for which they received abusive replies. Some are less visible, given that the period does cover a seven-month period.

\begin{table}[!htb]
\centering
\resizebox{0.85\textwidth}{!}{%
\begin{tabular}{|lrrrrr|}
\rowcolor[HTML]{EFEFEF} 
\hline
MP Name &
  \begin{tabular}[c]{@{}r@{}}Original\\ MP tweets\end{tabular} &
  \begin{tabular}[c]{@{}r@{}}Replies\\ to MP\end{tabular} &
  \begin{tabular}[c]{@{}r@{}}Abusive Replies\\ to MPs\end{tabular} &
  \begin{tabular}[c]{@{}r@{}}Sexist \\ Abuse\end{tabular} &
  \% Sexist \\ \hline \hline
Margaret Hodge      & 225   & 22,513  & 820    & 196 & 0.871 \\
Therese Coffey      & 169   & 25,520  & 1,558  & 169 & 0.662 \\
Margaret Ferrier    & 235   & 17,575  & 1,214  & 107 & 0.609 \\
Layla Moran         & 1,221 & 42,654  & 2,425  & 256 & 0.600 \\
Nadia Whittome      & 452   & 56,266  & 2,984  & 300 & 0.533 \\
Natalie Elphicke    & 80    & 12,290  & 609    & 65  & 0.529 \\
Naz Shah            & 220   & 19,115  & 1,248  & 97  & 0.507 \\
Theresa May         & 10    & 12,810  & 440    & 62  & 0.484 \\
Esther McVey        & 181   & 22,202  & 986    & 99  & 0.446 \\
Vicky Ford          & 245   & 12,810  & 694    & 52  & 0.406 \\
Dawn Butler         & 726   & 130,644 & 5,985  & 514 & 0.393 \\
Andrea Leadsom      & 427   & 21,987  & 794    & 85  & 0.387 \\
Zarah Sultana       & 621   & 103,519 & 3,869  & 388 & 0.375 \\
Andrea Jenkyns      & 210   & 64,860  & 3,797  & 241 & 0.372 \\
Nadine Dorries      & 324   & 117,348 & 4,854  & 428 & 0.365 \\
Diane Abbott        & 411   & 56,470  & 1,751  & 177 & 0.313 \\
Lucy Allan          & 655   & 64,662  & 2,532  & 201 & 0.311 \\
Barry Gardiner      & 109   & 18,664  & 3,257  & 56  & 0.300 \\
Selaine Saxby       & 744   & 20,691  & 1,720  & 59  & 0.285 \\
Imran Ahmad-Khan    & 245   & 29,996  & 2,786  & 84  & 0.280 \\
Ian Blackford       & 489   & 65,002  & 3,171  & 179 & 0.275 \\
Anneliese Dodds     & 390   & 22,695  & 489    & 59  & 0.260 \\
Caroline Lucas      & 960   & 57,248  & 1,304  & 144 & 0.252 \\
Rebecca Long-Bailey & 191   & 23,372  & 621    & 58  & 0.248 \\
Ed Davey            & 758   & 63,686  & 4,942  & 155 & 0.243 \\
Priti Patel         & 170   & 353,480 & 11,210 & 838 & 0.237 \\ \hline
\end{tabular}%
}
\caption{MPs receiving the most gendered abuse during the period studied from June - December 2020}
\label{tab:topSexistAbuse}
\end{table}

Figures \ref{fig:wc_male_abuse} and \ref{fig:wc_female_abuse} show word clouds for abuse directed at male and female MPs (respectively).

\begin{figure}
  \includegraphics[width=0.95\textwidth]{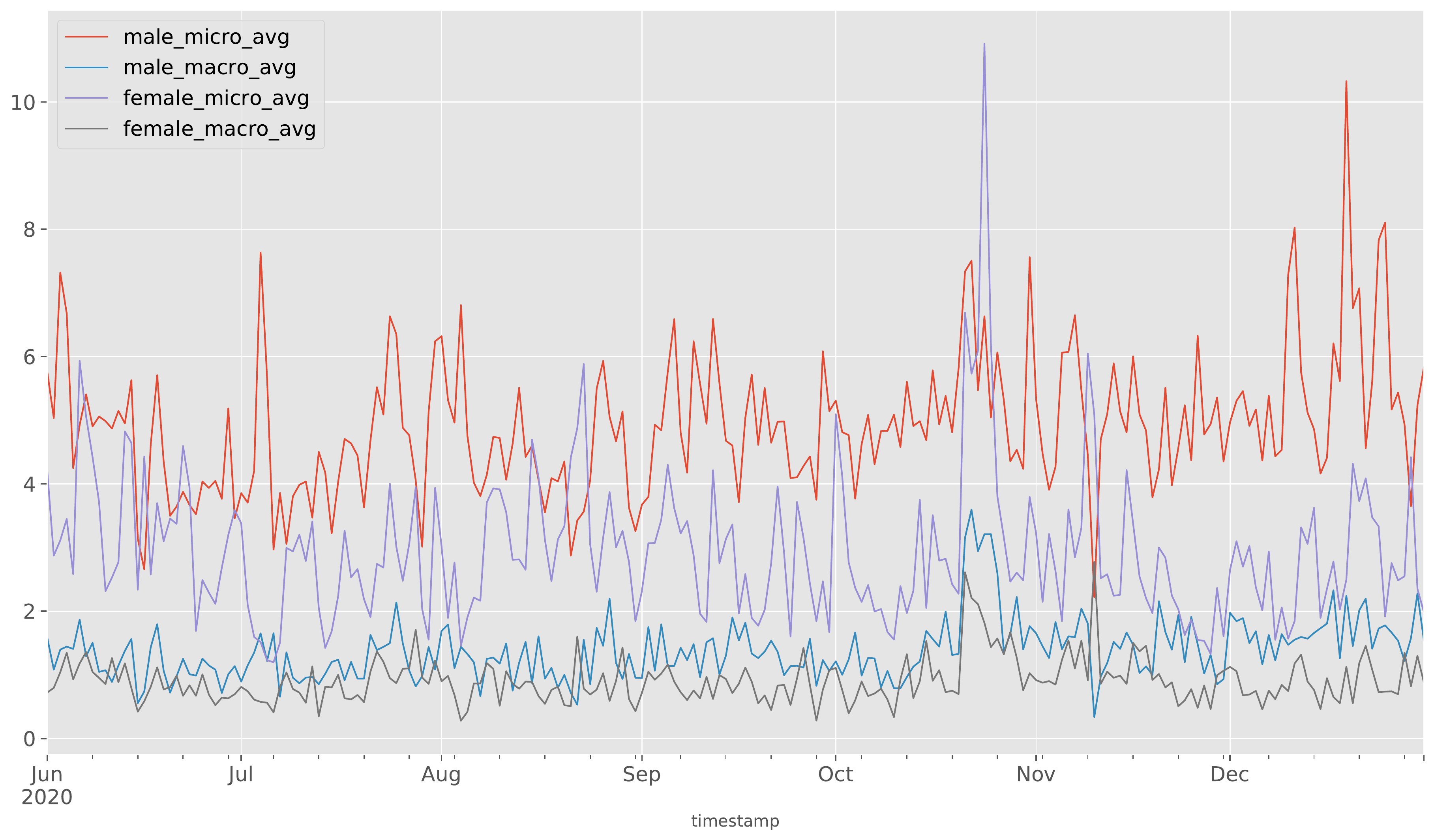}
  \caption{Percentage of abusive replies received by MPs of the respective gender.}
  \label{fig:tl_abuse_per_gender}
\end{figure}

Figure \ref{fig:tl_abuse_per_gender} shows the percentage of daily abusive replies per gender. Usually male MPs have a higher abusive tweet ratio (not accounting for the type of abusive language), but not always. For example, around 25 Oct, the micro-avg for female MPs is higher than for the male ones, despite the male macro-avg is still higher. Our analysis indicates that this was due to the conflict previously described between Chris Clarkson and Andrea Rayner. Also after the first week of Nov (around 8th), women MPs micro and macro average abuse exceeded those of men MPs. 

A more fine-grained analysis will be carried out in our planned follow-up report which will track online abuse towards MPs during one year of COVID-19 pandemic in the UK.

\section{Topical Hashtag Analysis}

\begin{figure}
  \includegraphics[width=.85\textwidth]{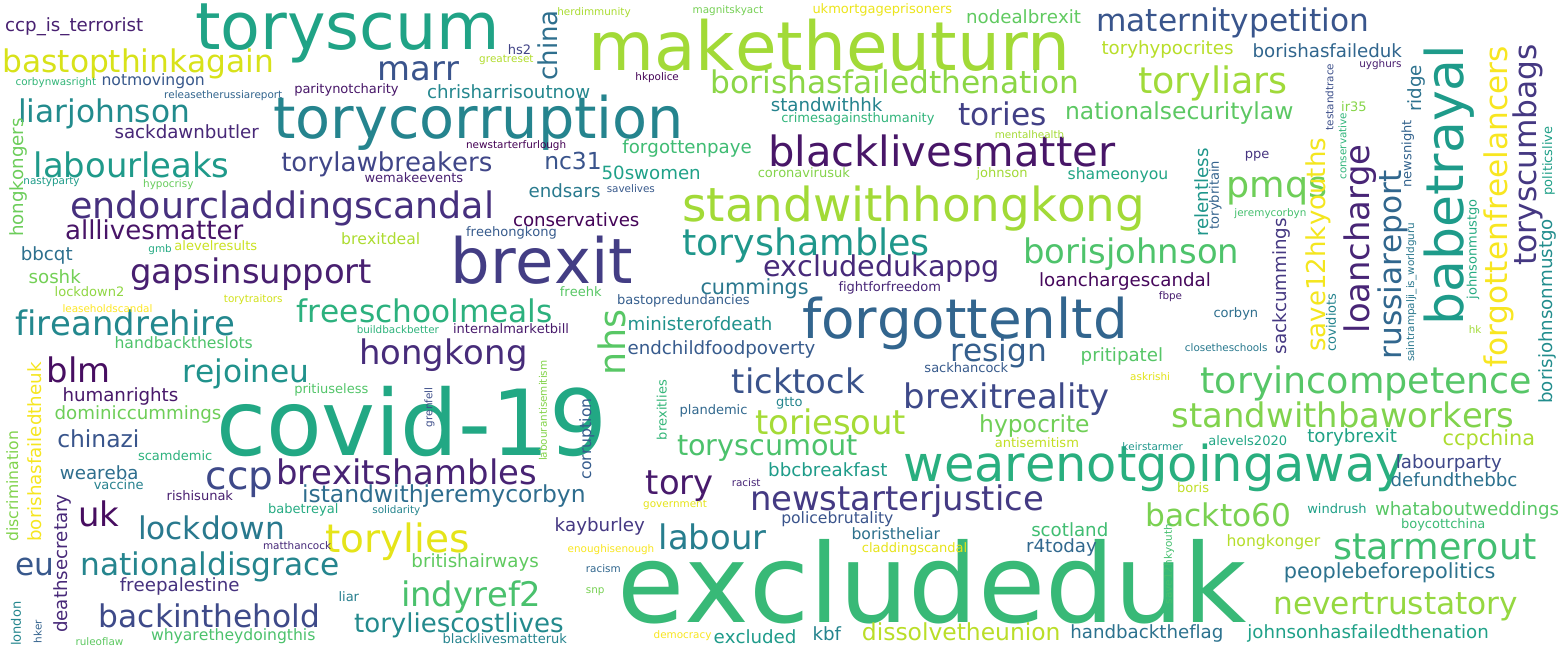}
  \caption{Hashtags appearing in replies to MPs, between June and December 2020.}
  \label{fig:wc_hts_allreplies}
\end{figure}

\begin{figure}
  \includegraphics[width=.85\textwidth]{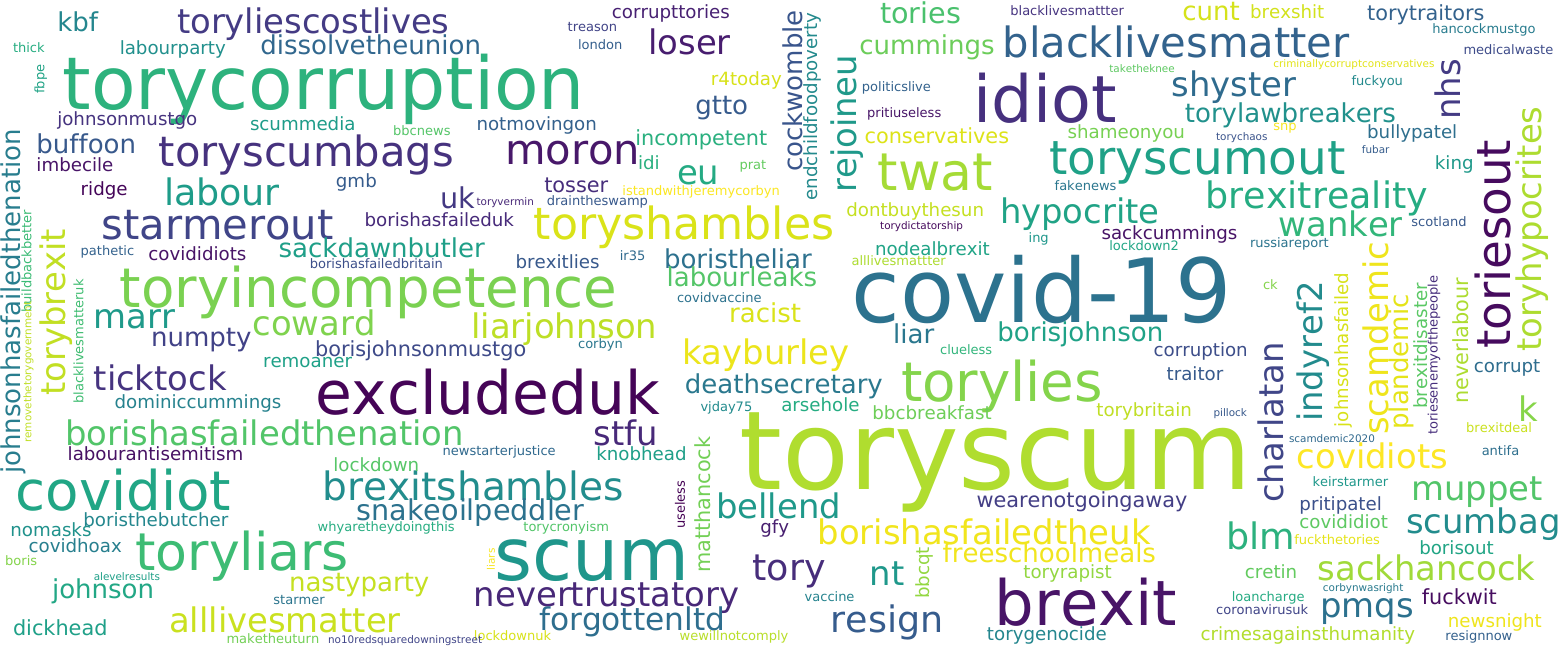}
  \caption{Hashtags appearing in abusive replies to MPs, between June and December 2020.}
  \label{fig:wc_hts_abusive_replies}
\end{figure}

In order to analyse the relationship between online abuse and topics such as Brexit, the governments' COVID-19 response and policies, and social issues, we conducted two types of analysis on hashtags: quantitative and qualitative. The first type of analysis is a frequency count of all hashtags observed in all replies during the period, and then by abusive replies only. We generated two word clouds (seen in Figures \ref{fig:wc_hts_allreplies} and \ref{fig:wc_hts_abusive_replies}) that represent hashtags used in all replies, and then in abusive replies, respectively. In Table \ref{tab:top15HTs}, we show the top 15 hashtags, and hashtags in the abusive tweets directed to MPs, from June to December 2020.

\begin{table}[!htb]
\centering
\resizebox{\textwidth}{!}{%
\begin{tabular}{|lr||lr|}
\hline
\rowcolor[HTML]{EFEFEF} 
\textbf{\begin{tabular}[c]{@{}l@{}}Hashtags in\\ all replies to MPs\end{tabular}} &
  \textbf{Count} &
  \textbf{\begin{tabular}[c]{@{}l@{}}Hashtags in abusive\\ replies to MPs\end{tabular}} &
  \textbf{Count} \\ \hline \hline
        excludeduk            & 76,006 & toryscum         & 2,250 \\
        covid-19              & 51,645 & covid-19         & 1,379 \\
        maketheuturn          & 26,102 & scum             & 917   \\
        toryscum              & 23,472 & torycorruption   & 889   \\
        brexit                & 20,863 & idiot            & 695   \\
        torycorruption        & 17,266 & brexit           & 667   \\
        ForgottenLtd          & 15,592 & excludeduk       & 592   \\
        BABetrayal            & 14,388 & toryincompetence & 474   \\
        wearenotgoingaway     & 12,897 & torylies         & 462   \\
        standwithhongkong     & 11,476 & covidiot         & 460   \\
        blacklivesmatter      & 9,367  & toryliars        & 432   \\
        ccp                   & 8,630  & twat             & 422   \\
        torylies              & 8,547  & toriesout        & 338   \\
        endourcladdingscandal & 7,884  & toryscumout      & 328   \\
        pmqs                  & 7,872  & toryshambles     & 305  \\ \hline
\end{tabular}%
}
\caption{Top 15 hashtags appearing in all replies, and in abusive replies to MPs, from June to December 2020.}
\label{tab:top15HTs}
\end{table}

\begin{table}[!htb]
\centering
\resizebox{\textwidth}{!}{
\begin{tabular}{|l|l|l|l|l|l|l|}
\hline
\rowcolor[HTML]{EFEFEF} 
\multicolumn{1}{|c|}{\cellcolor[HTML]{EFEFEF}\textbf{June}} &
  \multicolumn{1}{c|}{\cellcolor[HTML]{EFEFEF}\textbf{July}} &
  \multicolumn{1}{c|}{\cellcolor[HTML]{EFEFEF}\textbf{August}} &
  \multicolumn{1}{c|}{\cellcolor[HTML]{EFEFEF}\textbf{Sept}} &
  \multicolumn{1}{c|}{\cellcolor[HTML]{EFEFEF}\textbf{Oct}} &
  \multicolumn{1}{c|}{\cellcolor[HTML]{EFEFEF}\textbf{Nov}} &
  \multicolumn{1}{c|}{\cellcolor[HTML]{EFEFEF}\textbf{Dec}} \\ \hline
  
maketheuturn  & excludeduk  & excludeduk  & excludeduk  & toryscum  & excludeduk  & excludeduk \\ \hline
covid-19  & covid-19  & covid-19  & covid-19  & covid-19  & covid-19  & covid-19 \\ \hline
babetrayal  & \begin{tabular}[c]{@{}l@{}}wearenotgoinga\\ way\end{tabular}  & torycorruption  & brexit  & excludeduk  & toryscum  & brexit \\ \hline
excludedUK  & forgottenltd  & \begin{tabular}[c]{@{}l@{}}wearenotgoinga\\ way\end{tabular}  & \begin{tabular}[c]{@{}l@{}}wearenotgoinga\\ way\end{tabular}  & torycorruption  & torycorruption  & toryscum \\ \hline
\begin{tabular}[c]{@{}l@{}}blacklivesmatt\\ er\end{tabular}  & babetrayal  & brexit  & torycorruption  & freeschoolmeals  & brexit  & torylies \\ \hline
backinthehold  & \begin{tabular}[c]{@{}l@{}}standwithhong\\ kong\end{tabular}  & forgottenltd  & toryliars  & \begin{tabular}[c]{@{}l@{}}toryincompet\\ \hline ence\end{tabular}  & \begin{tabular}[c]{@{}l@{}}endourcladdi\\ ngscandal\end{tabular}  & \begin{tabular}[c]{@{}l@{}}brexitshamb\\ les\end{tabular} \\ \hline
\begin{tabular}[c]{@{}l@{}}standwithhong\\ kong\end{tabular}  & excludedukappg  & sackdawnbutler  & torylawbreakers  & toryscumbags  & forgottenltd  & torycorruption \\ \hline
forgottenltd  & ccp  & alevelresults  & toryshambles  & brexit  & \begin{tabular}[c]{@{}l@{}}peoplebefore\\ politics\end{tabular}  & forgottenltd \\ \hline
blm  & brexit  & toryshambles  & pmqs  & toryscumout  & starmerout  & \begin{tabular}[c]{@{}l@{}}toryincompet\\ ence\end{tabular} \\ \hline
\begin{tabular}[c]{@{}l@{}}bastopthinkag\\ ain\end{tabular}  & newstarterjustice  & alevels2020  & forgottenltd  & starmerout  & \begin{tabular}[c]{@{}l@{}}toryincompet\\ ence\end{tabular}  & brexitreality \\ \hline
\end{tabular}%
}
\caption{Top 10 hashtags per month from June - December 2020.}
\label{tab:hashtag_timeline}
\end{table}

As can be seen from these two tables of hashtags (Tables \ref{tab:top15HTs} and \ref{tab:hashtag_timeline}),  replies to MPs focused on a number of predominant themes, which can be categorised as: 
\begin{itemize}
    \item Dissatisfaction with government policies, e.g. tweets campaigning for the millions of people in the UK excluded from COVID-19 financial support (\#ExcludedUK; \#WeAreNotGoingAway);  appeals to the government to change its policies (\#MakeTheUTurn) on issues such as Brexit, free school meals, and COVID-19 financial support; the \#ForgottenLtd small businesses who aren't eligible for government support\footnote{\url{https://forgottenltd.com}}; the redundancies at British Airways\footnote{\url{https://twitter.com/search?q=\%23babetrayal}}; Brexit;
    \item Global political issues, e.g. Black Lives Matter, Hong Kong and China (\#StandingWithHongKong and \#CCP)
    \item Other issues, e.g. \#EndOurCladdingScandal (related to flammable cladding and the Grenfell tragedy);    
    \item Abusive hashtags aimed at the government, e.g. \#toryscum,  \#torycorruption, \#torylies.
\end{itemize}
When only abusive replies are considered, 12 of the top 15 dominating hashtags are critical of the Conservative Party and abusing the MPs (e.g. \#idiot, \#twat), with the other three being Covid-19, Brexit, and \#ExcludedUK.

For the second part of our analysis, we manually assessed a sub-set of 1,286 hashtags that were found more than 3 times in abusive replies to MPs during the period studied. We annotated each of these hashtags according to what type of issue the hashtag represents across the tweets where it was discovered. Table \ref{tab:ht_qualitative} provides a description of those codes, along with some examples of hashtags that belong to that category. 

\begin{figure}
  \includegraphics[width=.95\textwidth]{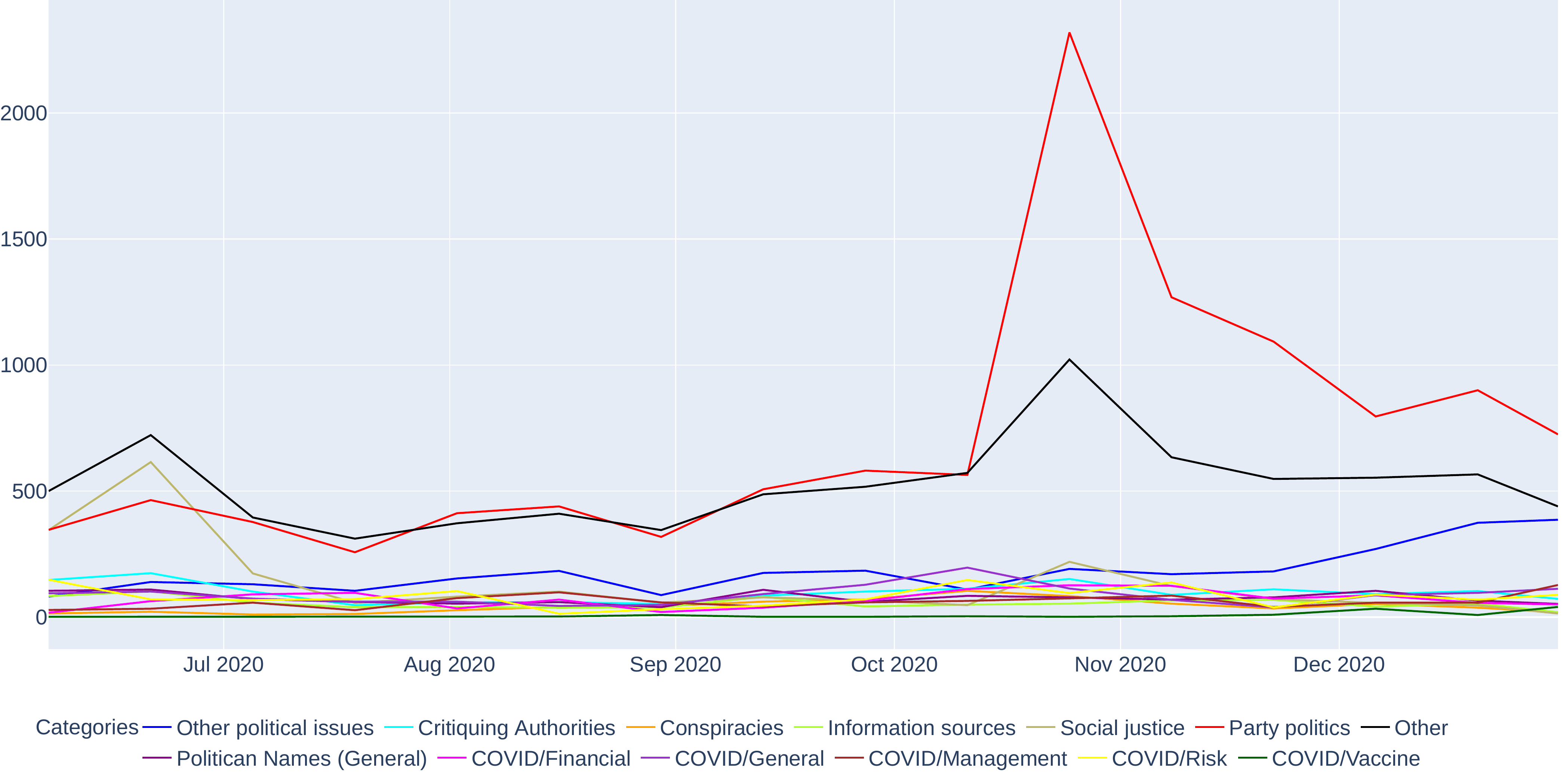}
  \caption{Timeline of hashtag categories and their volume  in abusive replies to MPs}
  \label{fig:info_seek_timeline}
\end{figure}

Our analysis in Figure \ref{fig:info_seek_timeline} shows that party politics have played a big role during the COVID-19 crisis, with many calls for different politicians to resign, or referring to specific scandals of each party. Other political issues, predominantly Brexit, also feature prominently.

\begin{table}[!htb]
\centering
\resizebox{\textwidth}{!}{%
\begin{tabular}{| >{\columncolor[HTML]{EFEFEF}}l ll|}
\hline
\textbf{Code} &
  \cellcolor[HTML]{EFEFEF}\textbf{Description} &
  \cellcolor[HTML]{EFEFEF}\textbf{Examples} \\ \hline\hline
\textbf{Critiquing Authorities} &
  \begin{tabular}[c]{@{}l@{}}Hashtags that are critical of the government \\ more generally, and it's activities\end{tabular} &
  \begin{tabular}[c]{@{}l@{}}\#cronyism, \#corruption, \\ \#lieslieslies\end{tabular} \\ \hline
\textbf{Information Sources} &
  \begin{tabular}[c]{@{}l@{}}Hashtags that are critiquing or reporting \\ about/from the media\end{tabular} &
  \begin{tabular}[c]{@{}l@{}}\#dontbuythesun, \#panorama, \\ \#buyapaper\end{tabular} \\ \hline
\textbf{Social Justice} &
  \begin{tabular}[c]{@{}l@{}}Hashtags that refer to issues of social \\ justice and equality, both in a positive and \\ negative sense\end{tabular} &
  \begin{tabular}[c]{@{}l@{}}\#IstandwithHongKong, \\ \#blacklivesmatter, \\ \#alllivesmatter\end{tabular} \\ \hline
\textbf{Politician Names} &
  \begin{tabular}[c]{@{}l@{}}Hashtags that refer to specific politicians \\ more neutrally\end{tabular} &
  \begin{tabular}[c]{@{}l@{}}\#BorisJohson, \#matthancock, \\ \#askRishi\end{tabular} \\ \hline
\textbf{Party Politics} &
  \begin{tabular}[c]{@{}l@{}}Hashtags that refer to specific parties, or\\ that refer to back and forth party politics\\ in the UK\end{tabular} &
  \begin{tabular}[c]{@{}l@{}}\#Toryscum, \#labourleaks, \\ \#80seatmajority\end{tabular} \\ \hline
\textbf{COVID- General} &
  General hashtags related to COVID-19 &
  \#Covid, \#corona, \#pandemic \\ \hline
\textbf{COVID- Financial} &
  \begin{tabular}[c]{@{}l@{}}COVID hashtags related to the financial\\ situation of citizens and the country\end{tabular} &
  \begin{tabular}[c]{@{}l@{}}\#excludedUK, \\ \#forgottenfreelancers,\\ \#forgottenltd\end{tabular} \\ \hline
\textbf{COVID- Risk} &
  \begin{tabular}[c]{@{}l@{}}COVID hashtags related to personal\\ risk and prevention\end{tabular} &
  \begin{tabular}[c]{@{}l@{}}\#wearamask, \\ \#socialdistancing, \#covidiot\end{tabular} \\ \hline
\textbf{COVID- Vaccine} &
  \begin{tabular}[c]{@{}l@{}}COVID hashtags related to vaccines and\\ the vaccine roll-out\end{tabular} &
  \begin{tabular}[c]{@{}l@{}}\#novaccine, \#coronavaccine, \\ \#covidvaccine\end{tabular} \\ \hline
\textbf{Conspiracies} &
  \begin{tabular}[c]{@{}l@{}}Hashtags related to conspiracies that have\\ arisen or been revived during COVID-19\end{tabular} &
  \begin{tabular}[c]{@{}l@{}}\#covidscam, \#fakecrisis, \\ \#scamdemic\end{tabular} \\ \hline
\textbf{Other political activities} &
  \begin{tabular}[c]{@{}l@{}}This code is for hashtags that relate to \\ other political activities that do not have \\ another code here.\end{tabular} &
  \begin{tabular}[c]{@{}l@{}}\#brexit, \#nodealcaroline, \\ \#chisbill\end{tabular} \\ \hline
\textbf{Other} &
  \begin{tabular}[c]{@{}l@{}}This code is for all other hashtags that \\ were too general (used in many different\\  circumstances) or ambiguous to code\end{tabular} &
  \begin{tabular}[c]{@{}l@{}}\#losers, \#muppets, \\ \#nationaldisgrace\end{tabular}\\\hline
\end{tabular}%
}
\caption{Codebook for hashtag qualitative analysis on }
\label{tab:ht_qualitative}
\end{table}

In June, we see several social justice issues take the forefront, possibly as a result of global racial justice protests. Hashtags about food security, workers rights, racial justice and global conflicts dominate this period (which we see as the olive green line peaks in June). Counter hashtags, such as alllivesmatter, appear toward the end of the list and a number of the hashtags associated with abusive language refer to social justice issues, indicating a small backlash to attention on social justice issues. What is conspicuously missing during this time, is prominence of the Brexit theme, despite just being 6 months away from the December 31st deadline. Some negative attention on the conservatives begins to bubble up, after a brief period of low (compared to the average) abuse levels at the beginning of 2020.

In July, we see the primary focus of attention shift to workers rights, and in particular those workers whose income has been disrupted by the pandemic, but for whom no provisions had (as of yet) been made. The exception to this shift away from more diverse social issues is the continued support for protests in Hong Kong. Brexit returns to the top of the hashtag themes. In terms of hashtags gathering abusive replies, negative attention on the conservatives grows in July. 

In August, a number of issues remain on the table from the previous months, including those forgotten by existing economic packages, Brexit, and negative attention on the conservatives. Concern about student A-levels arose during this period, after it was revealed that the algorithmic approach to deciding student grades (after cancelling 2020 exams) had resulted in marks below their teachers' predictions for more than a third of students\footnote{\url{https://en.wikipedia.org/wiki/2020_UK_GCSE_and_A-Level_grading_controversy}}. Another big issue during this time concerns a report leaked in April of 2020, which detailed anti-semitism and other types of racist and sexist abuse in the Labour party. Senior labour officials pushed back with claims that parts of the report had been falsified or taken out of context in early August. A push to encourage senior Labour officials to publish full reports, to clarify their comments, emerged in response. During that same week, Dawn Butler posted on Twitter that she had been racially profiled and stopped by the police on the road. Rumours that she had edited this video, or misrepresented what happened lead to attention and abuse.  
In September, anger and frustration with the conservative party is prominent as financial insecurity becomes the top concern that is visible in our analysis. While the crisis in Hong Kong is still receiving attention, concerns closer to home take priority. 

Attention to food security rises again in October, as the half-term time returns, the question of free school meals to the docket. Again, negative attention on the conservatives is still high.  The protests in Nigeria to  end the Special Anti-Robbery Squad (SARS), a police unit accused of multiple abuses, appear in our top ten list, in particular after the The Lekki Toll Gate Massacre on October 20th, 2020 (only recently re-opened in February 2021). 

In November, just 2 months before the Brexit deadline, attention on the topic rises again. Negative attention on the conservatives for multiple issues during this time, including continued dissatisfaction with job retention and financial schemes, the cladding scandal and pressures on the NHS, continue to occupy a top position. Nearly all of the hashtags associated with abusive content are about the conservatives, with the exception of a small amount of attention on Keir Starmer, potentially for criticising Boris Johnson heavily during this time. During Prime Minister's Questions, Starmer called Johnson the ``single biggest threat to the future of the UK''\footnote{\url{https://www.theguardian.com/politics/video/2020/nov/18/pmqs-keir-starmer-says-boris-johnson-is-single-biggest-threat-to-future-of-uk-video}}. 

By December, concern about Brexit is really beginning to take shape. Nine of the top hashtags are related to Brexit, four of which relate to abusive replies. Negative and abusive attention on the conservatives occupy the other significant portion of public attention during this time. Despite the continued lockdowns and the confusing government guidelines around Christmas holidays.

\section{Conspiracy theories}
Conspiracy thinking has been implicated in ``prejudice, witch hunts, revolutions, and genocide'' as well as terror attacks and rejection of scientific consensus. They are also a regular part of sense-making, in which people want to explain significant events that do not, as of yet, have a satisfactory explanation \cite{douglas2019understanding}. Conspiracy thinking has been shown to ``reduce normative political engagement'', while increasing ``non-normative political engagement'' \cite{RImhoff2021}. This is viewed as evidence of the connection between political extremism and violence. So, potentially lashing out at MPs with abusive language or threats could potentially be viewed as a non-normative political action.

In our last paper, we found more examples of conspiracy-related hashtags that involved the origins of the virus or it's connections to Chinese labs \cite{farrell2020vindication}. This is potentially due to the lack of clear information at the beginning of the pandemic. In our current period, top hashtags found in abusive replies to MPs (Table \ref{tab:th_conspiracy}) show that conspiracies related to covid being a scam or part of a larger plan to disrupt the freedoms of the people are at the top. This could potentially have to do with continued lockdowns and fears from various industries that have been unable to trade for a considerable time period. 

Some conspiracies existed before the COVID-19 crisis (kbf is actually Keep Britain Free, which is conspiracy adjacent), whereas others are more specific to COVID-19. We see in the narrative of tweets including these hashtags that older conspiracy theories remerge in the context of COVID-19, such as the great reset and conspiracies about Bill Gates. These topics return again and again and represent fears that the wealthy and powerful will seek to use their influence to control citizens. What's important to remember is that conspiracies often have a kernel of truth. There are many ways that those with money and influence can shape our experiences, which can be evidenced. However, there is no evidence of a large-scale, coordinated effort to control world population, engage in a cultural genocide or to implant micro-chips in our brains. 

In the second part of our analysis, we performed a small manual coding exercise on conspiracy hashtags to understand what those hashtags represent. The code ``pumping up the base'' had to do with any hashtag that is about communicating the presence of a movement of those interested in that hashtag, for example, the popular qanon slogan ``Where We Go One, We Go All'' (wwg1wga), or 3pointfivepercent, which refers to the so-called critical mass required make social change\footnote{\url{https://www.bbc.com/future/article/20190513-it-only-takes-35-of-people-to-change-the-world}}. Other conspiracies implicate the government or other authorities in wanting to control the people (such as Agenda21, or the GreatReset). This category seems to be the most highly represented. In the second largest categories are hashtags that communicate that the pandemic is not real (such as scamdemic, covidhoax). We have also coded a smaller category of hashtags that communicate that the pandemic is over-exaggerated (rather than completely fake, though individuals using this hashtag may believe that to be true). This includes hashtags like casedemic, which refer to potential anomalies in reporting on COVID-19 data.

\begin{table}[!htb]
\centering
\resizebox{0.75\textwidth}{!}{%
\begin{tabular}{|lcl|}
\hline
\rowcolor[HTML]{EFEFEF} 
    \textbf{Hashtag}      & \textbf{Hashtag Count} & \textbf{Hashtag Covid} \\\hline\hline
    scamdemic             & 115                    & Covid Conspiracy       \\
    kbf                   & 94                     & Existing Conspiracy    \\
    plandemic             & 77                     & Covid Conspiracy       \\
    covidhoax             & 43                     & Covid Conspiracy       \\
    scamdemic2020         & 28                     & Covid Conspiracy       \\
    coverup               & 27                     & Existing Conspiracy    \\
    thegreatreset         & 26                     & Existing Conspiracy    \\
    nwo                   & 25                     & Existing Conspiracy    \\
    covid1984             & 21                     & Covid Conspiracy       \\
    greatreset            & 20                     & Existing Conspiracy    \\
    3point5percent        & 18                     & Existing Conspiracy    \\
    casedemic             & 13                     & Covid Conspiracy       \\
    agenda21              & 12                     & Existing Conspiracy    \\
    wwg1wga               & 11                     & Existing Conspiracy    \\
    hoax                  & 10                     & Existing Conspiracy    \\
    fakepandemic          & 9                      & Covid Conspiracy       \\
    fakevirus             & 9                      & Covid Conspiracy       \\
    billgatesbioterrorist & 9                      & Existing Conspiracy    \\
    coronahoax            & 8                      & Covid Conspiracy       \\
    plandemichoax         & 8                      & Covid Conspiracy       \\
    billgates             & 8                      & Existing Conspiracy   \\ \hline
\end{tabular}%
}
\caption{Top conspiracy hashtags.}
\label{tab:th_conspiracy}
\end{table}

\begin{table}[!htb]
\centering
\resizebox{0.75\textwidth}{!}{%
\begin{tabular}{|>{\columncolor[HTML]{EFEFEF}}l |l|}
\hline
{\color[HTML]{000000} \textbf{Hashtag Category}}                 & \cellcolor[HTML]{EFEFEF}{\color[HTML]{000000} \textbf{Hashtag Count}} \\ \hline \hline
{\color[HTML]{000000} government/other authorities want control} & {\color[HTML]{000000} 368}                                            \\
{\color[HTML]{000000} pandemic is fake}             & {\color[HTML]{000000} 253}          \\
{\color[HTML]{000000} pandemic is over-exaggerated} & {\color[HTML]{000000} 22}           \\
{\color[HTML]{000000} pumping up the base}          & {\color[HTML]{000000} 37}           \\\hline
{\color[HTML]{000000} \textbf{Total}}               & {\color[HTML]{000000} \textbf{680}} \\ \hline
\end{tabular}%
}
\caption{Number of hashtags found in each hashtag category from June - December 2020.}
\label{tab:ht_conspiracy_categories}
\end{table}

\section{Conclusion}

In this paper, we have presented a follow-up work to our investigation of the first months of the pandemic, to provide an overview of trends in abuse toward UK MPs during the COVID-19 pandemic. We have presented quantitative analysis on the volume and frequency of abusive replies to UK MPs, as well as the various topical hashtags that are linked to those responses. We analysed the extent to which abuse levels appeared to be impacted by the features of party and gender, two unclear variables from previous research. We also included a deeper qualitative examination of the data, including a descriptive timeline of events that explain some of the levels of abuse we see at different times, directed toward different individuals. We also manually annotated and analysed how different clusters of hashtags appeared in the data, and looked more closely at the representation of conspiracies in those hashtags. 

Our analyses indicate that COVID-19 has added up to 1\% more abusive replies to levels that had  remained around 4\% over the previous four years. We found a clear party difference in this period, with the Conservatives receiving the lion's share of abusive replies. Our hashtag analysis also confirms these findings. 

While we were not able to detect considerable differences in the type of abusive speech that women and men MPs received, women were more likely to make our top list of MPs who receive gendered abuse. 

As with our previous work, we were able to confirm that prominence, personal characteristics and events do appear to make a difference in the amount of abusive replies a UK politicians will receive. 

\begin{backmatter}



\section*{Acknowledgements}
The authors thank Mark Greenwood for help with data extraction and Genevieve Gorrell for making available the abuse analysis code she used for analysing the initial COVID-19 period and previously -- abuse towards MPs in the run up to the 2019 election.

This research was partially supported by an ESRC grant ES/T012714/1 ``Responsible AI for Inclusive, Democratic Societies: A cross-disciplinary approach to detecting and countering abusive language online.'' and by an EC H2020 grant number 871042 (SoBigData++).  


\bibliographystyle{bmc-mathphys} 
\bibliography{covid-twitter-mp-abuse-white-paper}      

\end{backmatter}
\end{document}